\documentclass{article}

\usepackage[utf8]{inputenc}
\usepackage{amssymb}
\usepackage{amsmath}
\usepackage{authblk}
\usepackage{graphicx}
\usepackage[hyphens]{url}
\usepackage{hyperref}
\usepackage{enumitem}
\usepackage{algorithm}
\usepackage{booktabs}
\usepackage{bm}
\usepackage[toc,page]{appendix}		% Appendix numbering
\usepackage{subcaption} 			% Captions for subfigures
\usepackage{array}
\usepackage{setspace}
\usepackage{textcomp}
\newenvironment{conditions}[1][where]
{#1 \begin{tabular}[t]{>{$}l<{$} @{${}{}$} l}}
	{\end{tabular}\\[\belowdisplayskip]}

\begin{document}
\title{Decision-making with multiple correlated binary outcomes in clinical trials}

\author[1]{X.M. Kavelaars\thanks{E-mail: \texttt{x.m.kavelaars@tilburguniversity.edu}}}
\author[1,2]{J. Mulder}
\author[2]{M.C. Kaptein} 

\affil[1]{Department of Methodology and Statistics, Tilburg University, Tilburg, The Netherlands}
\affil[2]{Jheronimus Academy of Data Science, 's Hertogenbosch, The Netherlands}

\date{ }
\maketitle

%%%%%%%%%%%%%%%%%%%%%%%%%%%%%%%%%%%%%%%%%%%%%%%%%%%%%%%%%%%%%%%%%%%%%%%%%%%%%
%																			%
%								ABSTRACT 									%
%																			%
%%%%%%%%%%%%%%%%%%%%%%%%%%%%%%%%%%%%%%%%%%%%%%%%%%%%%%%%%%%%%%%%%%%%%%%%%%%%%	
\newpage
\begin{abstract}
	
Clinical trials often evaluate multiple outcome variables to form a comprehensive picture of the effects of a new treatment.
The resulting multidimensional insight contributes to clinically relevant and efficient decision-making about treatment superiority. 
Common statistical procedures to make these superiority decisions with multiple outcomes have two important shortcomings however: 1) Outcome variables are often modeled individually, and consequently fail to consider the relation between outcomes; and 2) superiority is often defined as a relevant difference on a single, on any, or on all outcomes(s); and lacks a compensatory mechanism that allows large positive effects on one or multiple outcome(s) to outweigh small negative effects on other outcomes.
To address these shortcomings, this paper proposes 1) a Bayesian model for the analysis of correlated binary outcomes based on the multivariate Bernoulli distribution; and 2) a flexible decision criterion with a compensatory mechanism that captures the relative importance of the outcomes.
A simulation study demonstrates that efficient and unbiased decisions can be made while Type I error rates are properly controlled. 
The performance of the framework is illustrated for 1) fixed, group sequential, and adaptive designs; and 2) non-informative and informative prior distributions.
\end{abstract}

%%%%%%%%%%%%%%%%%%%%%%%%%%%%%%%%%%%%%%%%%%%%%%%%%%%%%%%%%%%%%%%%%%%%%%%%%%%%%
%																			%
%								SECTION 1 									%
%																			%
%%%%%%%%%%%%%%%%%%%%%%%%%%%%%%%%%%%%%%%%%%%%%%%%%%%%%%%%%%%%%%%%%%%%%%%%%%%%%		
\newpage 
\section{Introduction}\label{sec:introduction}

	Clinical trials often aim to compare the effects of two treatments. 
	To ensure clinical relevance of these comparisons, trials are typically designed to form a comprehensive picture of the treatments by including multiple outcome variables.
	Collected data about efficacy (e.g. reduction of disease symptoms), safety (e.g. side effects), and other relevant aspects of new treatments are combined into a single, coherent decision regarding treatment superiority.
	An example of a trial with multiple outcomes is the CAR-B (Cognitive Outcome after WBRT or SRS in Patients with Brain Metastases) study, which investigated an experimental treatment for cancer patients with multiple metastatic brain tumors \cite{schimmel2018}.
	Historically, these patients have been treated with radiation of the whole brain (Whole Brain Radiation Therapy; WBRT).
	This treatment is known to damage healthy brain tissue and to increase the risk of (cognitive) side effects.
	More recently, local radiation of the individual metastases (stereotactic surgery; SRS) has been proposed as a promising alternative that saves healthy brain tissue and could therefore reduce side effects. 
	The CAR-B study compared these two treatments based on cognitive functioning, fatigue, and several other outcome variables \cite{schimmel2018}.

  	Statistical procedures to arrive at a superiority decision have two components: 1) A statistical model for the collected data; and 2) A decision rule to evaluate the treatment in terms of superiority based on the modelled data. 
  	Ideally, the combination of these components forms a decision procedure that satisfies two criteria: Decisions should be clinically relevant and efficient.
  	Clinical relevance ensures that the statistical decision rule corresponds to a meaningful superiority definition,	given the clinical context of the treatment.
 	Commonly used decision rules define superiority as one or multiple treatment difference(s) on the most important outcome, on any of the outcomes, or on all of the outcomes \cite{FDA2017, Murray2016,Sozu2012,Sozu2016}.
   	Efficiency refers to achieving acceptable error rates while minimizing the number of patients in the trial. 
  	The emphasis on efficiency is motivated by several considerations, such as small patient populations, ethical concerns, limited access to participants, and other difficulties to enroll a sufficient number of participants \cite{VandeSchoot2020}.
 	In the current paper, we address clinical relevance and efficiency in the context of multiple binary outcomes and propose a framework for statistical decision-making.

 	In trials with multiple outcomes, it is common to use a univariate modeling procedure for each individual outcome and combine these with one of the aforementioned decision rules \cite{FDA2017,Murray2016}.  	
 	Such decision procedures can be inefficient since they ignore the relationships between outcomes.
 	Incorporating these relations in the modeling procedure is crucial as they directly influence the amount of evidence for a treatment difference as well as the sample size required to achieve satisfactory error rates.
 	A multivariate modeling procedure takes relations between outcomes into account and can therefore be a more efficient and accurate alternative when outcomes are correlated.

 	Another interesting feature of multivariate models is that they facilitate the use of decision rules that combine multiple outcomes in a flexible way, for example via a compensatory mechanism.
   	Such a mechanism is characterized by the property that beneficial effects are given the opportunity to compensate adverse effects. 
   	The flexibility of compensatory decision-making is appealing, since a compensatory mechanism can be naturally extended with impact weights that explicitly take the clinical importance of individual outcome variables into account \cite{Murray2016}.
   	With impact weights, outcome variables of different importances can be combined into a single decision in a straightforward way.

\begin{figure}[htbp]
   		\centering
   		\includegraphics[width=0.4\textwidth,keepaspectratio]{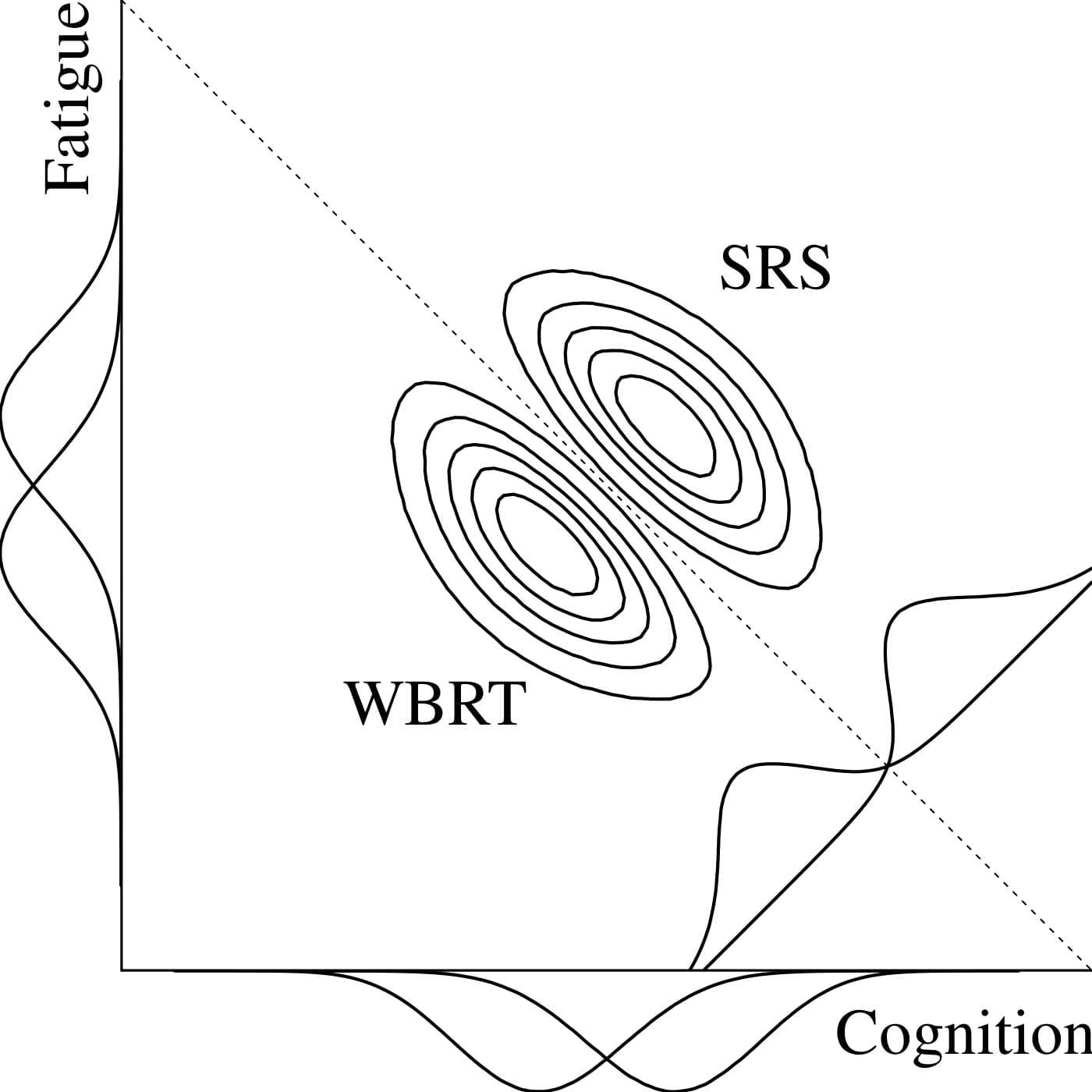}\\
   		\caption{Separation of two bivariate distributions (diagonally) versus separation of their univariate distributions (horizontally/vertically) for the CAR-B study. The dashed diagonal line represents a Compensatory decision rule with equal weights. Each distribution reflects the plausibility of the treatment effects on cognitive functioning and fatigue after observing fictive data.}
   		\label{fig:efficiency}
\end{figure}

   	Compensatory rules do not only contribute to clinical relevance, but also have the potential to increase trial efficiency.
   	Effects on individual outcomes may be small (and seemingly unimportant) while the combined treatment effect may be large (and important) \cite{OBrien1984,Tang1989,Pocock1987}, as visualized in Figure \ref{fig:efficiency} for fictive data of the CAR-B study. 
   	The two displayed bivariate distributions reflect the effects and their uncertainties on cognitive functioning and fatigue for SRS and WBRT.
   	The univariate distributions of both outcomes overlap too much to clearly distinguish the two treatments on individual outcome variables or a combination of them.
   	The bivariate distributions however clearly distinguish between the two treatments. 
   	Consequently, modeling a compensatory treatment effect with equal weights (visualized as the diagonal dashed line) would provide sufficient evidence to consider SRS superior in the presented situation.

 	In the current paper, we propose a decision procedure for multivariate decision-making with multiple (correlated) binary outcomes. 	
	The procedure consists of two components.
	First, we model the data with a multivariate Bernoulli distribution, which is a multivariate generalization of the univariate Bernoulli distribution. 
	The model is exact and does not rely on numerical approximations, making it appropriate for small samples.  
	Second, we extend multivariate analysis with a compensatory decision rule to include more comprehensive and flexible definitions of superiority.

	The decision procedure is based on a Bayesian multivariate Bernoulli model with a conjugate prior distribution.
	The motivation for this model is twofold.
	First, the multivariate Bernoulli model is a natural generalization of the univariate Bernoulli model, which intuitively parametrizes success probabilities per outcome variable. 
	Second, a conjugate prior distribution can greatly facilitate computational procedures for inference. 
	Conjugacy ensures that the form of the posterior distribution is known, making sampling from the posterior distribution straightforward.

	Although Bayesian analysis is well-known to allow for inclusion of information external to the trial by means of prior information \cite{Gelman2013},
	researchers who wish not to include prior information can obtain results similar to frequentist analysis. 
	The use of a non-informative prior distribution essentially results in a decision based on the likelihood of the data, such that 1) Bayesian and frequentist (point) estimates are equivalent; and 2) the frequentist p-value equals the Bayesian posterior probability of the null hypothesis in one-sided testing \cite{Marsman2017}.
	Since a (combined) p-value may be difficult to compute for the multivariate Bernoulli model, Bayesian computational procedures can exploit this equivalence and facilitate computations involved in Type I error control \cite{FDA2010,Wilson2019}.

	The remainder of the paper is structured as follows. 
	In the next section, we present a multivariate approach to the analysis of multiple binary outcomes. 
	Subsequently, we discuss various decision rules to evaluate treatment differences on multiple outcomes.
	The framework is evaluated in the \textit{\nameref{sec:evaluation}} section, and we discuss limitations and extensions in the \textit{\nameref{sec:discussion}}.
%
%
%%%%%%%%%%%%%%%%%%%%%%%%%%%%%%%%%%%%%%%%%%%%%%%%%%%%%%%%%%%%%%%%%%%%%%%%%%%%%
%																			%
%								SECTION 2 									%
%																			%
%%%%%%%%%%%%%%%%%%%%%%%%%%%%%%%%%%%%%%%%%%%%%%%%%%%%%%%%%%%%%%%%%%%%%%%%%%%%%	

\section{A model for multivariate analysis of multiple binary outcomes}\label{sec:analysis}
	\subsection{Notation}
	We start the introduction of our framework with some notation.
	The joint response for patient $i$ in treatment $j$ on $K$ outcomes will be denoted by $\bm{x}_{j,i}=(x_{j,i,1}, \dots, x_{j,i,K})$, where $i \in \{1,\dots, n_j\}$, and $j \in \{E,C\}$ (i.e., Experimental and Control).
	The response on outcome $k$ $x_{j,i,k} \in \{0,1\}$ ($0=$ failure, $1=$ success), such that $\bm{x}_{j,i}$ can take on $Q=2^{K}$ different combinations $\{1 \dots 11\}, \{1 \dots 10\}, \dots, \{0 \dots 01\}, \{0 \dots 00\}$.
	The observed frequencies of each possible response combination for treatment $j$ in a dataset of $n_{j}$ patients are denoted by vector $\bm{s}_{j}$ of length $Q$.
	The elements of $\bm{s}_{j}$ add up to $n_{j}$, $\sum_{q=1}^{Q} \bm{s}_{j,q}=n_{j}$.

	Vector $\bm{\theta}_{j}=(\theta_{j,1},\dots,\theta_{j,K})$ reflects success probabilities of $K$ outcomes for treatment $j$ in the population. 
	Vector $\bm{\delta}=(\delta_{1},\dots,\delta_{K})$ then denotes the treatment differences on $K$ outcomes, where  $\delta_{k}=\theta_{E,k}-\theta_{C,k}$.
	We use $\bm{\phi}_{j}=\phi_{j,1 \dots 11}, \phi_{j,1 \dots 10}, \dots, \phi_{j,0 \dots 01}, \phi_{j,0 \dots 00}$ to refer to probabilities of joint responses in the population, where $\phi_{j,q}$ denotes the probability of joint response combination $\bm{x}_{j,i}$ with configuration $q$. 
	Vector $\bm{\phi}_{j}$ has $Q$ elements, and sums to unity, $\sum_{q=1}^{Q} \bm{\phi}_{j,q}=1$.
	Information about the relation between outcomes $k$ and $l$ is reflected by $\phi_{j,kl}$, which is defined as the sum of those elements of $\bm{\phi}_{j}$ that have the $k^{th}$ and $l^{th}$ elements of $q$ equal to $1$, e.g. $\phi_{j,11}$ for $K=2$.
	Similarly, marginal probability $\theta_{j,k}$ follows from summing all elements of $\bm{\phi}_{j}$ with the $k^{th}$ element of $q$ equal to $1$. 
	For example, with three outcomes, the success probability of the first outcome is equal to $\theta_{j,1} = \phi_{j,111} + \phi_{j,110} + \phi_{j,101} + \phi_{j,100}$.

	\subsection{Likelihood}
	The likelihood of joint response $\bm{x}_{j,i}$ follows a $K$-variate Bernoulli distribution \cite{Dai2013}:
	\begin{flalign}\label{eq:mvbern}
		p(\bm{x}_{j,i}|\bm{\phi}_{j})=&
		\text{ multivariate  Bernoulli}(\bm{x}_{j,i}|\bm{\phi}_{j})&&\\\nonumber
		=&
		\phi_{j,1 \dots 11}^{x_{j,1} \times \dots \times x_{j,K}} 
		\phi_{j,1 \dots 10}^{x_{j,1} \times \dots \times x_{j,K-1} (1-x_{j,K})} 
		\times \dots \times&&\\\nonumber
		&\phi_{j,0\dots 01}^{(1-x_{j,1})\times \dots \times (1-x_{j,K-1})x_{j,K}} 
		\phi_{j,0\dots 00}^{(1-x_{j,1} \times \dots \times 1-x_{j,K})}.%
		&&
	\end{flalign}
	\noindent The multivariate Bernoulli distribution in Equation \ref{eq:mvbern} is a specific parametrization of the multinomial distribution. 
	The likelihood of $n_{j}$ joint responses summarized by cell frequencies in $\bm{s}_{j}$ follows a $Q$-variate multinomial distribution with parameters $\bm{\phi}_{j}$:%
	\begin{flalign}\label{eq:mvmult}
		p(\bm{s}_{j}|\bm{\phi}_{j})=& \text{ multinomial}(\bm{s}_{j}|\bm{\phi}_{j})&&\\\nonumber
		\propto &
		\phi_{j,1\dots 11}^{s_{j,1\dots 11}} 
		\phi_{j,1\dots 10}^{s_{j,1\dots 10}} 
		\times \dots \times
		\phi_{j,0\dots 01}^{s_{j,0\dots 01}}
		\phi_{j,0\dots 00}^{s_{j,0\dots 00}}.%
		&&
	\end{flalign}

	Conveniently, the multivariate Bernoulli distribution is consistent under marginalization.
	That is, marginalizing a $K-$variate Bernoulli distribution with respect to $p$ variables results in a ($K-p$)-variate Bernoulli distribution \cite{Dai2013}.	
	Hence, the univariate Bernoulli distribution is directly related and results from marginalizing ($K-1$) variables.

	The pairwise correlation between variables $x_{j,k}$ and $x_{j,l}$ is reflected by $\rho_{x_{j,k},x_{j,l}}$ \cite{Dai2013}:
	\begin{flalign}\label{eq:rho_bibern}
	\rho_{x_{j,k}x_{j,l}}=&
	\frac{\theta_{j,kl}-\theta_{j,k}\theta_{j,l}}
	{\sqrt{\theta_{j,k}(1-\theta_{j,k})\theta_{j,l}(1-\theta_{j,l})}}.%
	&&
		\end{flalign}
	\noindent 
	This correlation is over the full range, i.e. $-1 \leq \rho_{x_{j,k},x_{j,l}} \leq 1$ \cite{Olkin2015}.

	\subsection{Prior and posterior distribution}
	A natural choice to model prior information about response probabilities $\bm{\phi}_{j}$ is the Dirichlet distribution, since a Dirichlet prior and multinomial likelihood form a conjugate combination. 
	The $Q$-variate prior Dirichlet distribution has hyperparameters 
	$\bm{\alpha}_{j}^{0}=(
	\alpha_{j,11 \dots 11}^{0}, 
	\alpha_{j,11 \dots 10}^{0},
	\dots, 
	\alpha_{j,00 \dots 01}^{0},
	\alpha_{j,00 \dots 00}^{0} 
	)$:  
	\begin{flalign}\label{eq:dirichlet_prior}
	p(\bm{\phi}_{j})=&
	\text{ Dirichlet}(\bm{\phi}_{j}|\bm{\alpha}^{0}_{j})&&\\\nonumber
	\propto&
	\phi_{j,1\dots 11}^{\alpha^{0}_{j,1\dots 11}-1} 
	\phi_{j,1\dots 10}^{\alpha^{0}_{j,1\dots 10}-1} 
	\times	\dots \times
	\phi_{j,0\dots 01}^{\alpha^{0}_{j,0\dots 01}-1} 
	\phi_{j,0\dots 00}^{\alpha^{0}_{j,0\dots 00}-1},%
	&&
	\end{flalign}
	\noindent where each of the prior hyperparameters $\bm{\alpha}^{0}_{j}$ should be larger than zero to ensure a proper prior distribution.

	The posterior distribution of $\bm{\phi}_{j}$ results from multiplying the likelihood and the prior distribution and follows a Dirichlet distribution with parameters $\bm{\alpha}^{n}_{j}=\bm{\alpha}^{0}_{j}+\bm{s}_{j}$:
	\begin{flalign}\label{eq:dirichlet_posterior}
	p(\bm{\phi}_{j}|\bm{s}_{j}) = &
		\text{Dirichlet}(\bm{\phi}_{j}|\bm{\alpha}^{0}_{j} + \bm{s}_{j})&& \\\nonumber
		\propto &
	\phi_{j,1 \dots 11}^{s_{j,1 \dots 11}}
	\phi_{j,1 \dots 10}^{s_{j,1 \dots 10}}
	\times	\dots \times
	\phi_{j,0 \dots 01}^{s_{j,0 \dots 01}}
	\phi_{j,0 \dots 00}^{s_{j,0 \dots 00}} \times 
	&&\\\nonumber
	&
	\phi_{j,1 \dots 11}^{\alpha^{0}_{j,1 \dots 11}-1} 
	\phi_{j,1 \dots 10}^{\alpha^{0}_{j,1 \dots 10}-1}
	\times	\dots \times
	\phi_{j,0 \dots 01}^{\alpha^{0}_{j,0 \dots 01}-1} 
	\phi_{j,0 \dots 00}^{\alpha^{0}_{j,0 \dots 00}-1}
	&&\\\nonumber
	\propto&
	\phi_{j,1 \dots 11}^{\alpha^{n}_{j,1 \dots 11}-1} 
	\phi_{j,1 \dots 10}^{\alpha^{n}_{j,1 \dots 10}-1} 
	\times \dots \times
	\phi_{j,0 \dots 01}^{\alpha^{n}_{j,0 \dots 01}-1} 
	\phi_{j,0 \dots 00}^{\alpha^{n}_{j,0 \dots 00}-1}.
	&&
	\end{flalign}

	Since prior hyperparameters $\bm{\alpha}^{0}_{j}$ impact the posterior distribution of treatment difference $\bm{\delta}$, specifying them carefully is important. 
	Each of the hyperparameters contains information about one of the observed frequencies $\bm{s}_{j}$ and can be considered a prior frequency that reflects the strength of prior beliefs.	
	Equation \ref{eq:dirichlet_posterior} shows that the influence of prior information depends on prior frequencies $\bm{\alpha}^{0}_{j}$ relative to observed frequencies $\bm{s}_{j}$.
	When all elements of $\bm{\alpha}^{0}_{j}$ are set to zero, $\bm{\alpha}^{n}_{j} = \bm{s}_{j}$.
	This (improper) prior specification results in a posterior mean of $\phi_{j,q}|s_{j,q} = \frac{\alpha^{n}_{j,q}}{\sum_{p=1}^{Q} \alpha^{n}_{j,p}}$, which is equivalent to the frequentist maximum likelihood estimate of $\phi_{j,q} = \frac{s_{j,q}}{\sum_{p=1}^{Q} s_{j,p}}$.
	To take advantage of this property with a proper non-informative prior, one could specify hyperparameters slightly larger than zero such that the posterior distribution is essentially completely based on the information in the data (i.e. $\bm{\alpha}^{n}_{j} \approx \bm{s}_{j}$).

	To include prior information - when available - in the decision, $\bm{\alpha}^{0}_{j}$ can be set to specific prior frequencies to increase the influence on the decision.
	These prior frequencies may for example be based on results from related historical trials.
	We provide more technical details on prior specification in Appendix \textit{\nameref{app:prior}}.
	There we also highlight the relation between the Dirichlet distribution and the multivariate beta distribution,	and demonstrate that the prior and posterior distributions of $\bm{\theta}_{j}$ are multivariate beta distributions.

	The final superiority decision relies on the posterior distribution of treatment difference $\bm{\delta}$.
	Although this distribution does not belong to a known family of distributions, we can approach the distribution of $\bm{\delta}$ via a two-step transformation of the posterior samples of $\bm{\phi}_{j}$.
	First, a sample of $\bm{\phi}_{j}$ is drawn from its known Dirichlet distribution. 
	Next, these draws can be transformed to a sample of $\bm{\theta}_{j}$ using the property that joint response frequencies sum to the marginal probabilities.
	Finally, these samples from the posterior distributions of $\bm{\theta}_{E}$ and $\bm{\theta}_{C}$ can then be transformed to obtain the posterior distribution of joint treatment difference $\bm{\delta}$, by subtracting draws of $\bm{\theta}_{C}$ from draws of $\bm{\theta}_{E}$, i.e. $\bm{\delta}=\bm{\theta}_{E}-\bm{\theta}_{C}$.
	Algorithm \ref{alg:fixed} in Subsection \nameref{sec_sub:implementation} includes pseudocode with the steps required to obtain a sample from the posterior distribution of $\bm{\delta}$.

%%%%%%%%%%%%%%%%%%%%%%%%%%%%%%%%%%%%%%%%%%%%%%%%%%%%%%%%%%%%%%%%%%%%%%%%%%%%%
%																			%
%								SECTION 3 									%
%																			%
%%%%%%%%%%%%%%%%%%%%%%%%%%%%%%%%%%%%%%%%%%%%%%%%%%%%%%%%%%%%%%%%%%%%%%%%%%%%%	
\section{Decision rules for multiple binary outcomes}\label{sec:decision}

	The current section discusses how the model from the previous section can be used to make treatment superiority decisions.
	Treatment superiority is defined by the posterior mass in a specific subset of the multivariate parameter space of $\bm{\delta}=(\delta_{1},\dots, \delta_{K})$.
	The complete parameter space will be denoted by $\mathcal{S}\subset (-1,1)^{K}$, and the superiority space will be denoted by $\mathcal{S}_{Sup}\subset S$.
	Superiority is concluded when a sufficiently large part of the posterior distribution of $\bm{\delta}$ falls in superiority region $\mathcal{S}_{Sup}$: 
	\begin{flalign}\label{eq:criterion}
		P(\bm{\delta}\in \mathcal{S}_{sup}|\bm{s}_{E},\bm{s}_{C})>p_{cut}
	\end{flalign}%
	\noindent where $p_{cut}$ reflects the decision threshold to conclude superiority.
	The value of this threshold should be chosen to control the Type I error rate $\alpha$.

	\subsection{Four different decision rules}
	\begin{figure}[htbp]
		\centering
		\begin{subfigure}[c]{0.35\linewidth}
			\includegraphics[width=\linewidth,keepaspectratio]{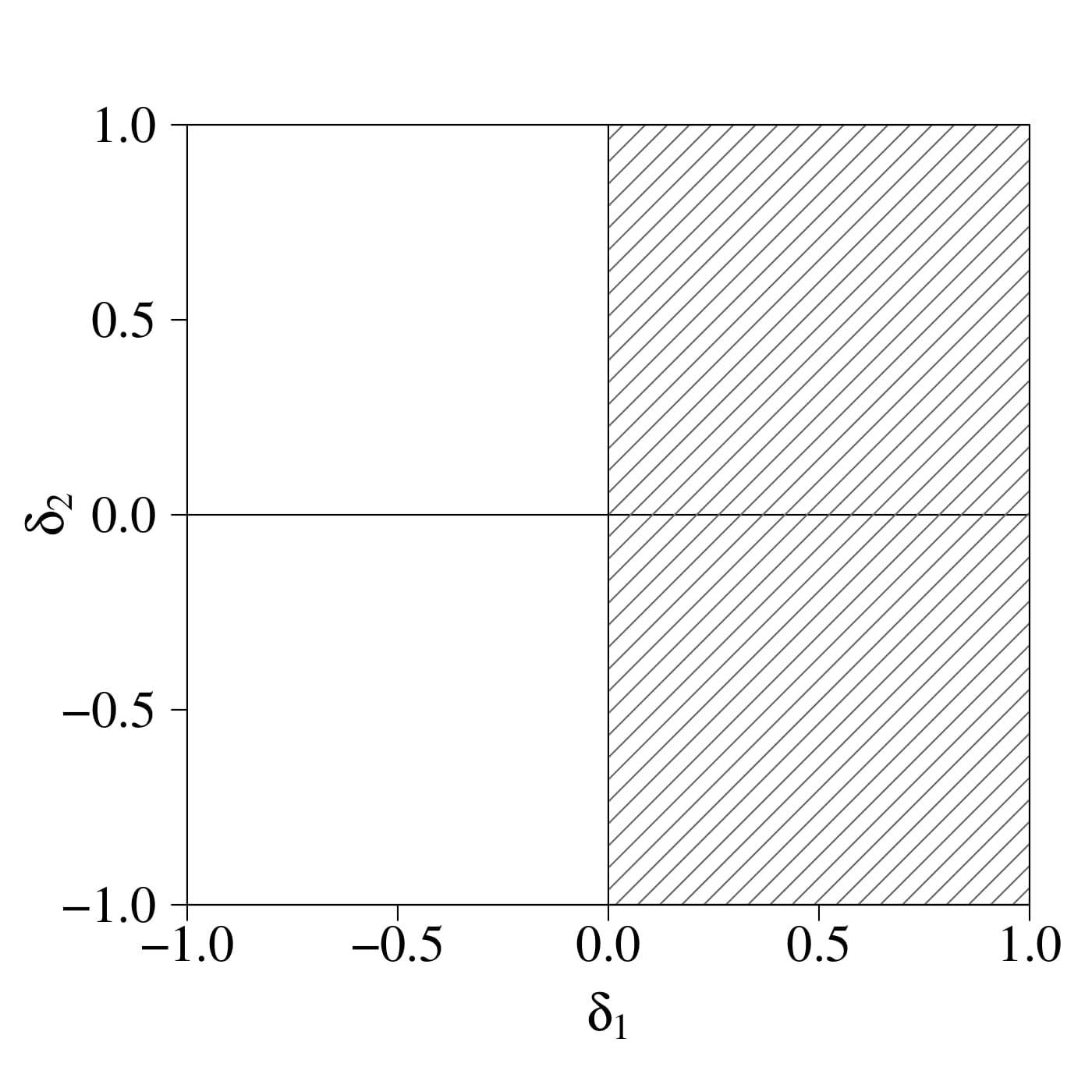}
			\caption{Single (outcome $1$)}\label{fig:sup_single}
		\end{subfigure}
		\begin{subfigure}[c]{0.35\linewidth}
			\includegraphics[width=\linewidth,keepaspectratio]{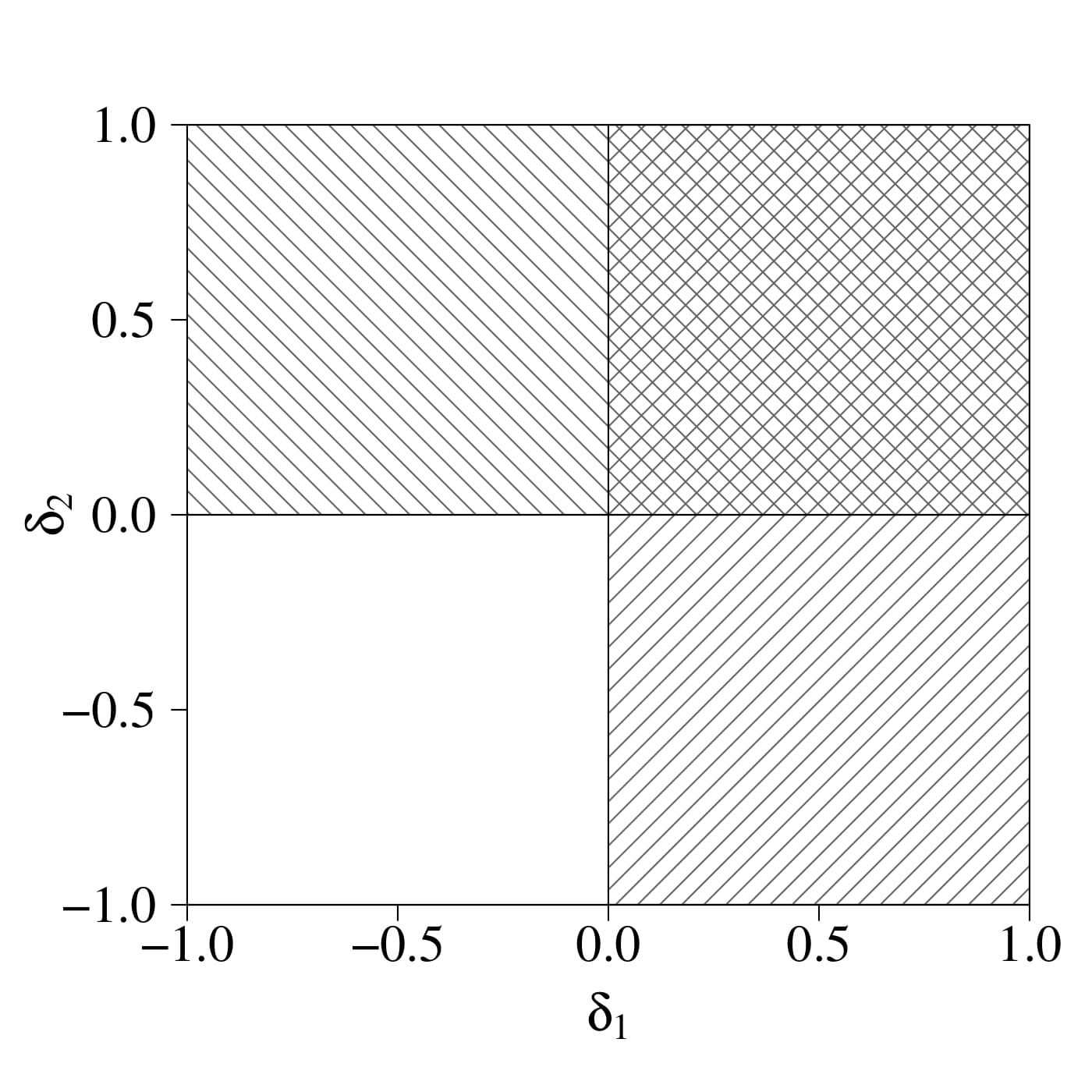}
			\caption{Any}\label{fig:sup_any}
		\end{subfigure}
		\begin{subfigure}[c]{0.35\linewidth}		
			\includegraphics[width=\linewidth,keepaspectratio]{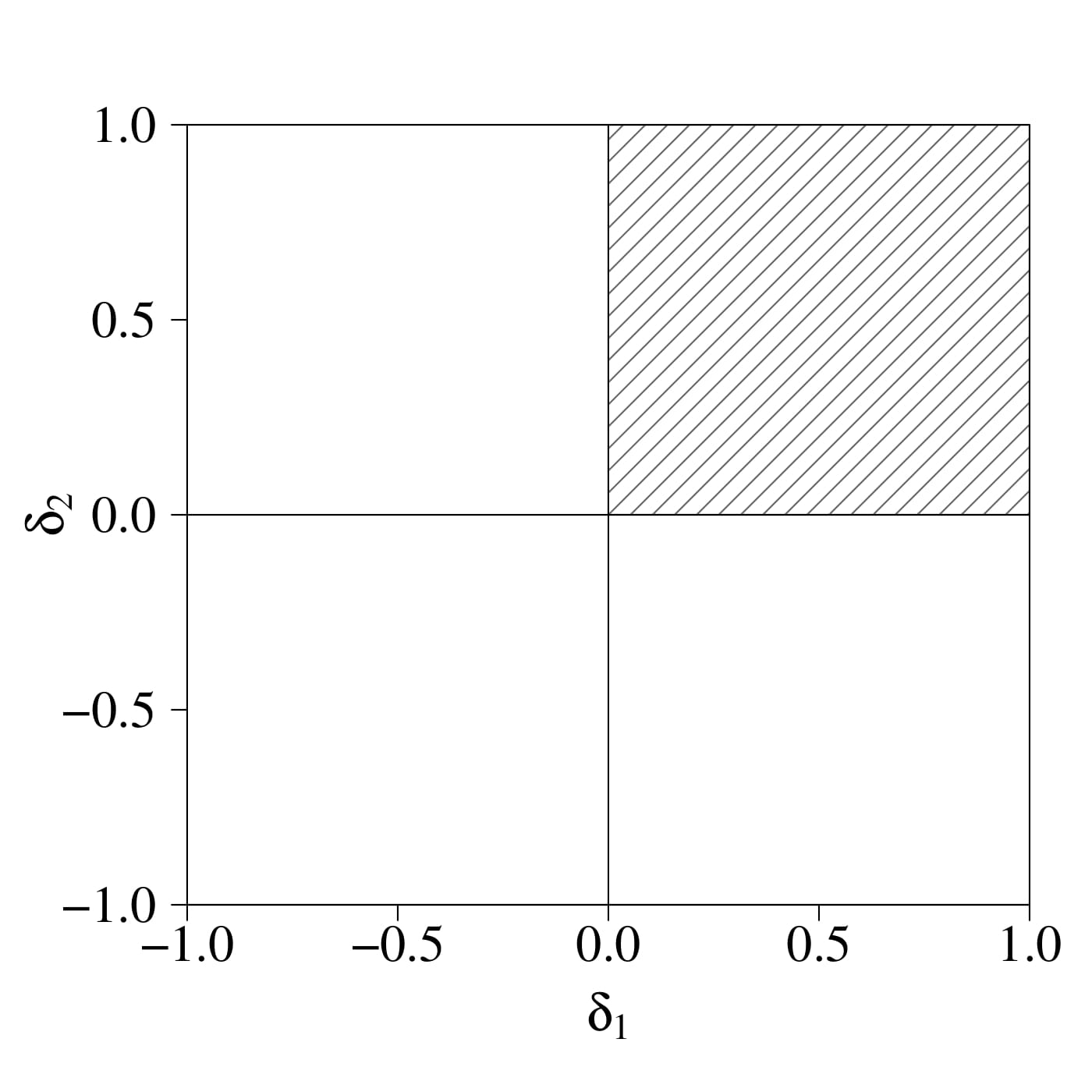}
			\caption{All}\label{fig:sup_all}
		\end{subfigure}
		\begin{subfigure}[c]{0.35\linewidth}		
			\includegraphics[width=\linewidth,keepaspectratio]{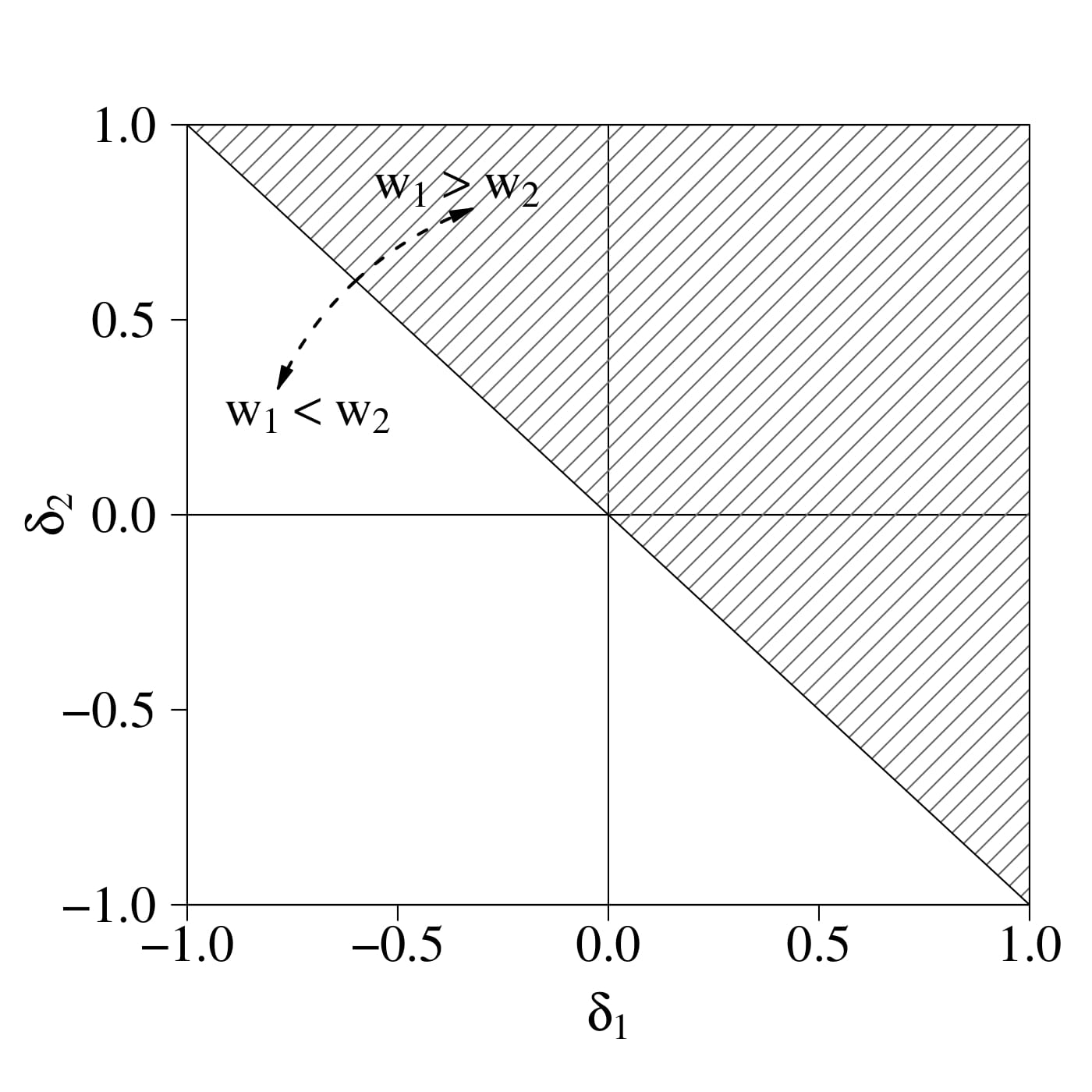}
			\caption{Compensatory}\label{fig:sup_compensatory}
		\end{subfigure}
		\caption{Superiority regions of various decision rules for two outcome variables ($K=2$). 
			The Any rule is a combination of the two Single rules.
			The Compensatory rule reflects $\bm{w}=(0.5,0.5)$.}
		\label{fig:superiority}
	\end{figure}

	Different partitions of the parameter space define different superiority criteria to distinguish two treatments. 
	The following decision rules conclude superiority when there is sufficient evidence that:
	\begin{enumerate}
		\item 
		\textit{Single rule:} 
		an a priori specified primary outcome $k$ has a treatment difference larger than zero. 
		The superiority region is denoted by:
		\begin{flalign}
			\mathcal{S}_{Single (k)}=\{\bm{\delta}| \delta_k>0 \}.
		\end{flalign}
   		\noindent
   		Superiority is concluded when 
   		\begin{flalign}
   		P(\bm{\delta} \in \mathcal{S}_{Single (k)}|\bm{s}_{E},\bm{s}_{C}) > p_{cut}.
   		\end{flalign}
   		\item 
		\textit{Any rule:} 
		at least one of the outcomes has a treatment difference larger than zero. 
		The superiority region is a combination of $K$ superiority regions of the Single rule: 
		\begin{flalign}
			\mathcal{S}_{Any}
			= & \{\mathcal{S}_{Single_{1}} \cup \dots \cup \mathcal{S}_{Single_{K}}\}. \nonumber
		\end{flalign}
		\noindent 
		Superiority is concluded when 
		\begin{flalign}
		\max_{k} P(\bm{\delta} \in \mathcal{S}_{Single (k)}|\bm{s}_{E},\bm{s}_{C}) > p_{cut}.
		\end{flalign}
		\item 
		\textit{All rule:}  
		all outcomes have a treatment difference larger than zero.
		Similar to the Any rule, the superiority region is a combination of $K$ superiority regions of the Single rule: 
		The superiority region is denoted by:
		\begin{flalign}
			\mathcal{S}_{All}= & 
			\{\mathcal{S}_{Single_{1}} \cap \dots \cap \mathcal{S}_{Single_{K}}\}.\nonumber
		\end{flalign}
	\noindent 
	Superiority is concluded when
	\begin{flalign}
	\min_{k} P(\bm{\delta} \in \mathcal{S}_{Single (k)}|\bm{s}_{E},\bm{s}_{C}) > p_{cut}.
	\end{flalign}
	\end{enumerate}
	\noindent 
	Next to facilitating these common decision rules, our framework allows for a Compensatory decision rule:
	\begin{enumerate}
		\setcounter{enumi}{3}
		\item 
		\textit{Compensatory rule:}
		the weighted sum of treatment differences is larger than zero. 
		The superiority region is denoted by:
		\begin{flalign} 
			\mathcal{S}_{Compensatory}(\bm{w})=\{\bm{\delta}| \sum_{k=1}^{K} w_{k}\delta_{k}>0\} 
		\end{flalign}
		\begin{conditions}
		\bm{w} & $=(w_{1},\dots,w_{K})$ reflect the weights for outcomes $1,\dots,K$,\\
		 0 & $\leq w_k \leq 1$ and $\sum_{k=1}^K w_{k}=1$.\\
		\end{conditions}
	\noindent 
	Superiority is then concluded when:
	\begin{flalign}
	P(\bm{\delta} \in \mathcal{S}_{Compensatory}(\bm{w})|\bm{s}_{E},\bm{s}_{C}) > p_{cut}.
	\end{flalign}
	\end{enumerate}
	\noindent Figure \ref{fig:superiority} visualizes these four decision rules.

	From our discussion of the different decision rules, a number of relationships between them can be identified.
	First, mathematically the Single rule can be considered a special case of the Compensatory rule with weight $w_{k}=1$ for primary outcome $k$ and $w_{l}=0$ for all other outcomes. 
	Second, the superiority region of the All rule is a subset of the superiority regions of the other rules, i.e. 
	\begin{flalign}
	\mathcal{S}_{All} \subset \mathcal{S}_{Single}, \mathcal{S}_{Compensatory}, \mathcal{S}_{Any}.
\end{flalign}
	The Single rule is in turn a subset of the superiority region of the Any rule, such that 
	\begin{flalign}
	\mathcal{S}_{Single} \subset \mathcal{S}_{Any}.
	\end{flalign}
	These properties can be observed in Figure \ref{fig:superiority} and translate directly to the amount of evidence provided by data $\bm{s}_{E}$ and $\bm{s}_{C}$. The posterior probability of the All rule is always smallest, while the posterior probability of the Any rule is at least as large as the posterior probability of the Single rule: 
	\begin{flalign}\label{eq:compare_psup}
		P(\mathcal{S}_{Any}|\bm{s}_{E},\bm{s}_{C}) \geq 
		P(\mathcal{S}_{Single}|\bm{s}_{E},\bm{s}_{C}) > 
		P(\mathcal{S}_{All}|\bm{s}_{E},\bm{s}_{C})&&&\\\nonumber
		P(\mathcal{S}_{Compensatory}|\bm{s}_{E},\bm{s}_{C}) >
		P(\mathcal{S}_{All}|\bm{s}_{E},\bm{s}_{C}).&&&
	\end{flalign}
	\noindent 
	The ordering of the posterior probabilities of different decision rules (Equation \ref{eq:compare_psup}) implies that superiority decisions are most conservative under the All rule and most liberal under the Any rule.
	In practice, this difference has two consequences. 
	First, to properly control Type I error probabilities for these different decision rules, one needs to set a larger decision threshold $p_{cut}$ for the Any rule than for the All rule. 
	Second, the All rule typically requires the largest sample size to obtain sufficient evidence for a superiority decision.

\begin{figure}[htbp]
		\centering
		\includegraphics[width=\textwidth,keepaspectratio]{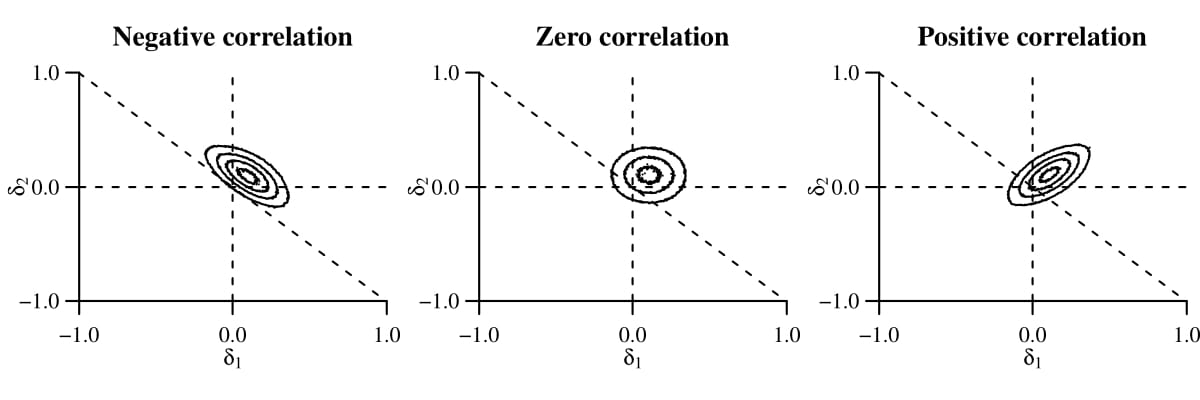}\\
		\caption{Influence of the correlation between two treatment differences on the proportion of overlap between the bivariate distribution of treatment differences $\bm{\delta}$ and the superiority regions.}
		\label{fig:correlation}
	\end{figure}

	Additionally, the correlation between treatment differences, $\rho_{\delta_{k},\delta_{l}}$, influences the posterior probability to conclude superiority.
	The correlation influences the overlap with the superiority region, as visualized in Figure \ref{fig:correlation}.
	Consequently, the Single rule is not sensitive to the correlation.
	A negative correlation requires a smaller sample size than a positive correlation under the Any and Compensatory rules, and vice versa for the All rule.

	\subsection{Specification of weights of the Compensatory decision rule}\label{sec:sub_weights}
	To utilize the flexibility of the Compensatory rule, researchers may wish to specify weights $\bm{w}$.
	The current subsection discusses two ways to choose these weights: Specification can be based on the impact of outcome variables or on efficiency of the decision.

	Specification of impact weights is guided by substantive considerations to reflect the relative importance of outcomes.
	When $\bm{w}=(\frac{1}{K},\dots,\frac{1}{K})$, all outcomes are equally important and all success probabilities in $\bm{\theta}_{j}$ exert an identical influence on the weighted success probability. 
	Any other specification of $\bm{w}$ that satisfies $\sum_{k=1}^{K}w_{k}=1$ implies unequal importance of outcomes.
	To make the implications of importance weight specification more concrete, let us reconsider the two potential side effects of brain cancer treatment in the CAR-B study: cognitive functioning and fatigue \cite{schimmel2018}. 
	When setting $(w_{cognition},w_{fatigue})=(0.50,0.50)$, both outcomes would be considered equally important and a decrease of (say) $0.10$ in fatigue could be compensated by an increase on cognitive functioning of at least $0.10$. 
	When $w_{cognition}>0.50$, cognitive functioning is more influential than fatigue; and vice versa when $w_{cognition}<0.50$.
	If $w_{cognition}=0.75$ and $w_{fatigue}=0.25$ for example, the treatment difference of cognitive functioning has three times as much impact on the decision as the treatment difference of fatigue.

	Efficiency weights are specified with the aim of optimizing the required sample size.
	As the weights directly affect the amount of evidence for a treatment difference, the efficiency of the Compensatory decision rule can be optimized with values of $\bm{w}$ that are a priori expected to maximize the probability of falling in the superiority region.
	This strategy could be used when efficiency is of major concern, while researchers do not have a strong preference for the substantive priority of specific outcomes. 
	The technical details required to find efficient weights are presented in Appendix \textit{\nameref{app:weights}}.

\subsection{Implementation of the framework}\label{sec_sub:implementation}
The procedure to arrive at a decision using the multivariate analysis procedure proposed in the previous sections is presented in Algorithm \ref{alg:fixed} for a design with fixed sample size $n_{j}$ of treatment $j$.
We present the algorithm for designs with interim analyses in Algorithm \ref{alg:interim} in Appendix \textit{\nameref{app:implementation_adaptive}}.

\begin{algorithm}
	\caption{Decision procedure for a fixed design}
	\label{alg:fixed}
	\begin{tabular}{p{\textwidth}}
		\begin{enumerate}[label={\arabic*},leftmargin=7pt,topsep=0pt,parsep=0pt,labelindent=0pt,itemindent=0pt,listparindent=-5pt]
			\item 
			\underline{\textbf{Initialize}}		
			
			\begin{enumerate}[label={},leftmargin=7pt,topsep=0pt,parsep=0pt,labelindent=0pt,itemindent=0pt,listparindent=-5pt]
				\item %while\for
				\begin{enumerate}[label={\alph*},leftmargin=7pt,topsep=0pt,parsep=0pt,labelindent=0pt,itemindent=0pt,listparindent=-5pt]
					\item 
					Choose decision rule
					
					\begin{enumerate}[label={},leftmargin=7pt,topsep=0pt,parsep=0pt,labelindent=0pt,itemindent=0pt,listparindent=-5pt]
						\item 
						\textbf{if} Compensatory
						\textbf{then} specify weights $\bm{w}$	
						
						\item 
						\textbf{if} Single
						\textbf{then} specify $k$	
						
						\item 
						\textbf{end if}
					\end{enumerate}
					
				\end{enumerate}
				
				\item
				\textbf{for} each treatment $j \in	\{E,C\}$
				\textbf{do}
				
				\begin{enumerate}[label={\alph*},leftmargin=7pt,topsep=0pt,parsep=0pt,labelindent=0pt,itemindent=0pt,listparindent=-5pt]
					\setcounter{enumiii}{1}
					\item 
					Choose prior hyperparameters $\bm{\alpha}^{0}_{j}$
					
				\end{enumerate}
				\item 
				\textbf{end for}
				
				\begin{enumerate}[label={\alph*},leftmargin=7pt,topsep=0pt,parsep=0pt,labelindent=0pt,itemindent=0pt,listparindent=-5pt]
					\setcounter{enumiii}{2}
					\item 
					Choose Type I error rate $\alpha$ and power $1-\beta$ 
					
					\item 
					Determine decision threshold $p_{cut}$
									
					\begin{enumerate}[label={},leftmargin=7pt,topsep=0pt,parsep=0pt,labelindent=0pt,itemindent=0pt,listparindent=-5pt]
						\item 
						\textbf{if} Any rule
						\textbf{then} $1- \frac{1}{2} \alpha$
						
						\item 
						\textbf{else} $1-\alpha$
							
						\item 
						\textbf{end if}
					\end{enumerate}
					
					\item
					Determine sample size $n_{j}$ based on anticipated treatment differences $\bm{\delta}^{n}$
				\end{enumerate}
			\end{enumerate}
			\item 
			\underline{\textbf{Collect data and compute evidence}}%	
			\begin{enumerate}[label={},leftmargin=7pt,topsep=0pt,parsep=0pt,labelindent=0pt,itemindent=0pt,listparindent=-5pt]
				\item 
				\textbf{for} each treatment $j \in \{E,C\}$
				
				\begin{enumerate}[label={\alph*},leftmargin=7pt,topsep=0pt,parsep=0pt,labelindent=0pt,itemindent=0pt,listparindent=-5pt]
					\item
					Collect $n_{j}$ joint responses $\bm{x}_{j,i}$
					
					\item
					Compute joint response frequencies $\bm{s}_{j}$
					
					\item
					Compute posterior parameters $\bm{\alpha}^{n}_{j} = \bm{s}_{j} + \bm{\alpha}^{0}_{j}$
					
					\item 
					Sample $L$ posterior draws, $\bm{\phi}^{l}_{j}$, $\bm{\phi}_{j}|\bm{\alpha}^{n}_{j} \sim Dirichlet(\bm{\phi}_{j}|\bm{\alpha}^{n}_{j})$
					
					\item 
					Sum draws $\bm{\phi}^{l}_{j}$ to $\bm{\theta}^{l}_{j}$
					
				\end{enumerate}
				\item 
				\textbf{end for}
				
				\begin{enumerate}[label={\alph*},leftmargin=7pt,topsep=0pt,parsep=0pt,labelindent=0pt,itemindent=0pt,listparindent=-5pt]
					\setcounter{enumiii}{5}
					\item 
					Transform draws $\bm{\theta}^{l}_{j}$ to $\bm{\delta}^{l}$ via $\delta^{l}_{k} = \theta^{l}_{E,k} - \theta^{l}_{C,k}$
					
					\item 
					Compute posterior probability of treatment superiority $P(\bm{\delta} \in \mathcal{S}_{Sup}|\bm{s}_{E},\bm{s}_{C})$ as the proportion of posterior draws in superiority region $\mathcal{S}_{Sup}$
					
				\end{enumerate}
			\end{enumerate}
			\item \underline{\textbf{Make final decision}}
			
			\begin{enumerate}[label={},leftmargin=7pt,topsep=0pt,parsep=0pt,labelindent=0pt,itemindent=0pt,listparindent=-5pt]
				\item 
				\textbf{if} $P(\bm{\delta} \in \mathcal{S}_{Sup}|\bm{s}_{E},\bm{s}_{C}) > P_{cut}$
				\textbf{then} conclude superiority
				
				\item 
				\textbf{else} conclude non-superiority
				
				\item 
				\textbf{end if}
			\end{enumerate}
		\end{enumerate}\\
	\end{tabular}
\end{algorithm}

%%%%%%%%%%%%%%%%%%%%%%%%%%%%%%%%%%%%%%%%%%%%%%%%%%%%%%%%%%%%%%%%%%%%%%%%%%%%%
%																			%
%								SECTION 4 									%
%																			%
%%%%%%%%%%%%%%%%%%%%%%%%%%%%%%%%%%%%%%%%%%%%%%%%%%%%%%%%%%%%%%%%%%%%%%%%%%%%%	
\section{Numerical evaluation}\label{sec:evaluation}

The current section evaluates the performance of the presented multivariate decision framework by means of simulation in the context of two outcomes ($K=2$).
We seek to demonstrate 1) how often the decision procedure results in an (in)correct superiority conclusion to learn about decision error rates; 2) how many observations are required to conclude superiority with satisfactory error rates to investigate the efficiency of different decision rules, and 3) how well the average estimated treatment difference corresponds to the true treatment difference to examine bias.
The current section is structured as follows.
We first introduce the simulation conditions, the procedure to compute sample sizes for each of these conditions, and the procedure to generate and evaluate data.
We then discuss the results of the simulation.

\paragraph{Conditions}
The performance of the framework is examined as a function of the following factors:
\begin{enumerate}
	\item 
	\textit{Data generating mechanisms:}
	We generated data of eight treatment difference combinations $\bm{\delta}^{T}$ and three correlations between outcomes $\rho_{\theta_{j,1},\theta_{j,2}}$. 
	An overview of these $8 \times 3 = 24$ data generating mechanisms is given in Table \ref{tab:conditions}.
	In the remainder of this section, we refer to these data generating mechanisms with numbered combinations (e.g. $1.2$), where the first number reflects treatment difference $\bm{\delta}^{T}$ and the second number refers to correlation $\rho_{\theta_{j,1}, \theta_{j,2}}^{T}$.
	
	\item 
	\textit{Decision rules:}
	The generated data were evaluated with six different decision rules. 
	We used the Single (for outcome $k=1$), Any, and All rules, as well as three different Compensatory rules: One with equal weights $\bm{w}=(0.50,0.50)$ and two with unequal weights $\bm{w}=(0.76,0.24)$ and $\bm{w}=(0.64,0.36)$.
	The weight combinations of the latter two Compensatory rules optimize the efficiency of data generating mechanisms with uncorrelated (i.e. $8.2$) and correlated (i.e. $8.1$) treatment differences respectively, following the procedure in Appendix \textit{\nameref{app:weights}}.
	We refer to these three Compensatory rules as Compensatory-Equal (C-E), Compensatory-Unequal Uncorrelated (C-UU) and Compensatory-Unequal Correlated (C-UC) respectively.
\end{enumerate}

% latex table generated in R 3.5.3 by xtable 1.8-3 package
% Mon Feb 03 11:27:09 2020
\begin{table}[htbp]
	\small\sf\centering
	\caption{Data generating mechanisms (DGM) used in numerical evaluation of the framework.} 
	\label{tab:conditions}
	\begin{tabular}{lrrrrrrrrr}
		\toprule
		DGM & $\delta_{1}^{T}$ & $\delta_{2}^{T}$ & $\rho_{\theta_{j,1}, \theta_{j,2}}^{T}$ & 
		$\theta_{E,1}^{T}$ & $\theta_{E,2}^{T}$ & $\phi_{E,11}^{T}$ &
		$\theta_{C,1}^{T}$ & $\theta_{C,2}^{T}$ & $\phi_{C,11}^{T}$ \\
		\midrule
		1.1 & -0.20 & -0.20 & -0.30 & 0.40 & 0.40 & 0.09 & 0.60 & 0.60 & 0.29 \\ 
		1.2 &   &   & 0.00 &   &   & 0.16 &   &   & 0.36 \\ 
		1.3 &   &   & 0.30 &   &   & 0.23 &   &   & 0.43 \\ 
		\multicolumn{10}{c}{ }\\
		2.1 & 0.00 & 0.00 & -0.30 & 0.50 & 0.50 & 0.17 & 0.50 & 0.50 & 0.17 \\ 
		2.2 &   &   & 0.00 &   &   & 0.25 &   &   & 0.25 \\ 
		2.3 &   &   & 0.30 &   &   & 0.32 &   &   & 0.32 \\ 
		\multicolumn{10}{c}{ }\\
		3.1 & 0.10 & 0.10 & -0.30 & 0.55 & 0.55 & 0.23 & 0.45 & 0.45 & 0.13 \\ 
		3.2 &   &   & 0.00 &   &   & 0.30 &   &   & 0.20 \\ 
		3.3 &   &   & 0.30 &   &   & 0.38 &   &   & 0.28 \\ 
		\multicolumn{10}{c}{ }\\
		4.1 & 0.20 & 0.20 & -0.30 & 0.60 & 0.60 & 0.29 & 0.40 & 0.40 & 0.09 \\ 
		4.2 &   &   & 0.00 &   &   & 0.36 &   &   & 0.16 \\ 
		4.3 &   &   & 0.30 &   &   & 0.43 &   &   & 0.23 \\ 
		\multicolumn{10}{c}{ }\\
		5.1 & 0.40 & 0.40 & -0.30 & 0.70 & 0.70 & 0.43 & 0.30 & 0.30 & 0.03 \\ 
		5.2 &   &   & 0.00 &   &   & 0.49 &   &   & 0.09 \\ 
		5.3 &   &   & 0.30 &   &   & 0.55 &   &   & 0.15 \\ 
		\multicolumn{10}{c}{ }\\
		6.1 & 0.40 & 0.00 & -0.30 & 0.70 & 0.50 & 0.28 & 0.30 & 0.50 & 0.08 \\ 
		6.2 &   &   & 0.00 &   &   & 0.35 &   &   & 0.15 \\ 
		6.3 &   &   & 0.30 &   &   & 0.42 &   &   & 0.22 \\ 
		\multicolumn{10}{c}{ }\\
		7.1 & 0.20 & -0.40 & -0.30 & 0.60 & 0.30 & 0.11 & 0.40 & 0.70 & 0.21 \\ 
		7.2 &   &   & 0.00 &   &   & 0.18 &   &   & 0.28 \\ 
		7.3 &   &   & 0.30 &   &   & 0.25 &   &   & 0.35 \\ 
		\multicolumn{10}{c}{ }\\
		8.1 & 0.24 & 0.08 & -0.30 & 0.62 & 0.54 & 0.26 & 0.38 & 0.46 & 0.10 \\ 
		8.2 &   &   & 0.00 &   &   & 0.33 &   &   & 0.17 \\ 
		8.3 &   &   & 0.30 &   &   & 0.41 &   &   & 0.25 \\ 
		\bottomrule
	\end{tabular}
\end{table}

\paragraph{Sample size computations}\label{sec:compute_n}
To properly control Type I error and power, each of the $24 \times 6$ conditions requires a specific sample size. 
These sample sizes $n_{j}$ are based on anticipated treatment differences $\bm{\delta}^{n}$, that corresponded to the true parameters of each data generating mechanism in Table \ref{tab:conditions} (i.e. $\bm{\delta}^{n} = \bm{\delta}^{T}$ and $\rho_{\theta_{j,1},\theta_{j,2}}^{n}=\rho_{\theta_{j,1},\theta_{j,2}}^{T}$).
Procedures to compute sample sizes per treatment group for the different decision rules were the following:
\begin{enumerate}
	\item 
	For the Single rule, we used a two-proportion $z-$test, where we plugged in the anticipated treatment difference on the first outcome variable (i.e $\delta_{1}^{n}$). 
	
	\item 
	Following Sozu et al. \cite{Sozu2010,Sozu2016} we used multivariate normal approximations of correlated binary outcomes for the All and Any rules.
	
	\item 
	For the Compensatory rule, we used a continuous normal approximation with mean $\sum_{k=1}^{K} w_{k} \theta_{j,k}$ and variance $\sum_{k=1}^{K} w^{2}_{k}\sigma^{2}_{j,k}+2\mathop{\sum\sum}\limits_{k < l} w_{k} w_{l} \sigma_{j,kl}$.
	Here, $\sigma^{2}_{j,k} = \theta_{j,k} (1 - \theta_{j,k})$ and $\sigma_{j,kl} = \phi_{j,kl} - \theta_{j,k} \theta_{j,l}$. 

\end{enumerate} 

The computed sample sizes are presented in Table \ref{tab:CompareRules_nStop}. 
Conditions that should not result in superiority were evaluated at sample size $n_{j}=1,000$.

\paragraph{Data generation and evaluation}
Of each data generating mechanism presented in Table \ref{tab:conditions}, we generated $5,000$ samples of size $2 \times n_{j}$. 
These data were combined with a proper uninformative prior distribution with hyperparameters $\bm{\alpha}^{0}_{j}=(0.01,\dots,0.01)$ to satisfy $\bm{\alpha}^{n}_{j} \approx \bm{s}_{j}$, as discussed in Section \nameref{sec:analysis}.
We aimed for Type I error rate $\alpha=.05$ and power $1-\beta=.80$, which corresponds to a decision threshold $p_{cut}$ of $1-\alpha=0.95$ (Single, Compensatory, All rules) and $1-\frac{1}{2}\alpha=0.975$ (Any rule) \cite{Sozu2012,Sozu2016,Marsman2017}. 
The generated datasets were evaluated using the procedure in steps $2$ and $3$ of Algorithm \ref{alg:fixed}.

The proportion of samples that concluded superiority reflects Type I error rates (when false) and power (when correct).
We assessed the Type I error rate under the data generating mechanism with the least favorable population values of $\bm{\delta}^{T}$ under the null hypothesis in frequentist one-sided significance testing.
These are values of $\bm{\delta}^{T}$ outside $\mathcal{S}_{Sup}$ that are most difficult to distinguish from values of $\bm{\delta}^{T}$ inside $\mathcal{S}_{Sup}$.
Adequate Type I error rates for the least favorable treatment differences imply that the Type I error rates of all values of $\bm{\delta}^{T}$ outside $\mathcal{S}_{Sup}$ are properly controlled.
The least favorable values of $\bm{\delta}^{T}$ were reflected by treatment difference $2$ for the Single, Any, and Compensatory rules, and treatment difference $6$ for the All rule.
Bias was computed as the difference between the observed treatment difference at sample size $n_{j}$ and the true treatment difference $\bm{\delta}^{T}$.

\subsection{Results}

% latex table generated in R 3.5.3 by xtable 1.8-3 package
% Mon Feb 17 08:01:58 2020
\begin{table}[htbp]
	\centering
	\caption{P(Conclude superiority) for different data generating mechanisms (DGM) and decision rules.
	Bold-faced values indicate the conditions with least favorable values.} 
	\label{tab:CompareRules_pSup}
	\begin{tabular}{lrrrrrr}
		\toprule
		DGM &
		\multicolumn{1}{l}{Single} & 
		\multicolumn{1}{l}{Any} &
		\multicolumn{1}{l}{All} & 
		\multicolumn{1}{l}{C-E} & 
		\multicolumn{1}{l}{C-UU}&
		\multicolumn{1}{l}{C-UC}\\
		\midrule
		1.1 & 0.000 & 0.000 & 0.000 & 0.000 & 0.000 & 0.000 \\ 
		1.2 & 0.000 & 0.000 & 0.000 & 0.000 & 0.000 & 0.000 \\ 
		1.3 & 0.000 & 0.000 & 0.000 & 0.000 & 0.000 & 0.000 \\ 
		\multicolumn{7}{c}{ }\\
		2.1 & \textbf{0.051} & \textbf{0.048} & 0.000 & 0.049 & 0.052 & 0.051 \\ 
		2.2 & 0.046 & 0.045 & 0.003 & 0.056 & 0.048 & 0.054 \\ 
		2.3 & 0.051 & 0.045 & 0.008 & \textbf{0.049} & \textbf{0.049} & \textbf{0.049} \\ 
		\multicolumn{7}{c}{ }\\
		3.1 & 0.810 & 0.796 & 0.801 & 0.807 & 0.804 & 0.790 \\ 
		3.2 & 0.799 & 0.801 & 0.804 & 0.806 & 0.788 & 0.791 \\ 
		3.3 & 0.799 & 0.807 & 0.809 & 0.800 & 0.797 & 0.803 \\ 
		\multicolumn{7}{c}{ }\\
		4.1 & 0.794 & 0.784 & 0.806 & 0.811 & 0.789 & 0.784 \\ 
		4.2 & 0.808 & 0.802 & 0.814 & 0.813 & 0.804 & 0.803 \\ 
		4.3 & 0.804 & 0.801 & 0.816 & 0.804 & 0.796 & 0.800 \\ 
		\multicolumn{7}{c}{ }\\
		5.1 & 0.807 & 0.806 & 0.830 & 0.881 & 0.817 & 0.857 \\ 
		5.2 & 0.807 & 0.814 & 0.838 & 0.831 & 0.813 & 0.813 \\ 
		5.3 & 0.809 & 0.847 & 0.822 & 0.809 & 0.798 & 0.802 \\ 
		\multicolumn{7}{c}{ }\\
		6.1 & 0.811 & 0.779 & 0.053 & 0.824 & 0.798 & 0.819 \\ 
		6.2 & 0.813 & 0.777 & 0.045 & 0.805 & 0.808 & 0.820 \\ 
		6.3 & 0.803 & 0.758 & \textbf{0.051} & 0.801 & 0.788 & 0.803 \\ 
		\multicolumn{7}{c}{ }\\
		7.1 & 0.799 & 0.789 & 0.000 & 0.000 & 0.863 & 0.002 \\ 
		7.2 & 0.804 & 0.792 & 0.000 & 0.000 & 0.857 & 0.003 \\ 
		7.3 & 0.807 & 0.794 & 0.000 & 0.000 & 0.867 & 0.005 \\ 
		\multicolumn{7}{c}{ }\\
		8.1 & 0.787 & 0.782 & 0.789 & 0.808 & 0.804 & 0.805 \\ 
		8.2 & 0.777 & 0.797 & 0.807 & 0.804 & 0.799 & 0.804 \\ 
		8.3 & 0.785 & 0.811 & 0.807 & 0.805 & 0.805 & 0.806 \\ 
		\bottomrule
	\end{tabular}
\end{table}

% latex table generated in R 3.5.3 by xtable 1.8-3 package
% Mon Feb 17 07:38:39 2020
\begin{table}[htbp]
	\small\sf\centering
	\caption{Average sample size to correctly conclude superiority for different data generating mechanisms (DGM) and decision rules.
	Bold-faced values indicate the lowest sample size per data generating mechanism.
Conditions with a hyphen should not result in treatment superiority.} 
	\label{tab:CompareRules_nStop}
	\begin{tabular}{lrrrrrr}
		\toprule
		DGM &  
		\multicolumn{1}{l}{Single} & 
		\multicolumn{1}{l}{Any} &
		\multicolumn{1}{l}{All} & 
		\multicolumn{1}{l}{C-E} & 
		\multicolumn{1}{l}{C-UU}&
		\multicolumn{1}{l}{C-UC}\\
		\midrule
		1.1 & - & - & - & - & - & - \\ 
		1.2 & - & - & - & - & - & - \\ 
		1.3 & - & - & - & - & - & - \\ 
		\multicolumn{7}{l}{ }\\
		2.1 & - & - & - & - & - & - \\ 
		2.2 & - & - & - & - & - & - \\ 
		2.3 & - & - & - & - & - & - \\ 
		\multicolumn{7}{l}{ }\\
		3.1 & 307 & 191 & 424 & \textbf{108} & 157 & 119 \\ 
		3.2 & 307 & 217 & 418 & \textbf{154} & 192 & 162 \\ 
		3.3 & 307 & 247 & 406 & \textbf{199} & 226 & 206 \\ 
		\multicolumn{7}{l}{ }\\
		4.1 & 75 & 47 & 105 & \textbf{26} & 39 & 29 \\ 
		4.2 & 75 & 53 & 103 & \textbf{38} & 47 & 40 \\ 
		4.3 & 75 & 60 & 101 & \textbf{49} & 55 & 50 \\ 
		\multicolumn{7}{l}{ }\\
		5.1 & 17 & 11 & 25 & \textbf{6} & 9 & 7 \\ 
		5.2 & 17 & 12 & 25 & \textbf{9} & 11 & \textbf{9} \\ 
		5.3 & 17 & 14 & 24 & \textbf{11} & 12 & \textbf{11} \\ 
		\multicolumn{7}{l}{ }\\
		6.1 & 17 & 21 & - & 25 & \textbf{15} & 17 \\ 
		6.2 & \textbf{17} & 21 & - & 36 & 19 & 24 \\ 
		6.3 & \textbf{17} & 21 & - & 47 & 22 & 30 \\ 
		\multicolumn{7}{l}{ }\\
		7.1 & \textbf{75} & 95 & - & - & 608 & - \\ 
		7.2 & \textbf{75} & 95 & - & - & 733 & - \\ 
		7.3 & \textbf{75} & 95 & - & - & 858 & - \\ 
		\multicolumn{7}{l}{ }\\
		8.1 & 51 & 56 & 482 & 41 & 38 & \textbf{36} \\ 
		8.2 & 51 & 60 & 482 & 59 & \textbf{46} & 49 \\ 
		8.3 & \textbf{51} & 63 & 482 & 76 & 55 & 62 \\ 
		\bottomrule
	\end{tabular}
\end{table}

The proportion of samples that concluded superiority and the required sample size are presented in Tables \ref{tab:CompareRules_pSup} and \ref{tab:CompareRules_nStop} respectively.
Type I error rates were properly controlled around $\alpha=.05$ for each decision rule under its least favorable data generating mechanism. 
The power was around $.80$ in all scenarios with true superiority.
Moreover, average treatment differences were estimated without bias (smaller than $0.01$ in all conditions).

Given these satisfactory error rates, a comparison of sample sizes provides insight in the efficiency of the approach. 
We remark here that a comparison of sample sizes is only relevant when the decision rules under consideration have a meaningful definition of superiority.
Further, in this discussion of results we primarily focus on the newly introduced Compensatory rule in comparison to the other decision rules. 
The results demonstrate that the Compensatory rule consistently requires fewer observations than the All rule, and often - in particular when treatment differences are equal (i.e. treatment differences $3-5$) - than the Any and the Single rule. 
Similarly, the Any rule consistently requires fewer observations than the All rule and could be considered an attractive option in terms of sample sizes.
Note however that the more lenient Any rule may not result in a meaningful decision for all trials, since the rule would also conclude superiority when the treatment has a small positive treatment effect and large negative treatment effect (i.e. treatment difference $7$); A scenario that may not be clinically relevant.

The influence of the relation between outcomes is also apparent: Negative correlations require fewer observations than positive correlations. 
The variation due to the correlation is considerable: The average sample size almost doubles in scenarios with equal treatment differences (e.g. data generating mechanisms $3.1$ vs. $3.3$ and $4.1$ vs. $4.3$).

Comparison of the three different Compensatory rules further highlights the influence of weights $\bm{w}$ and illustrates that a Compensatory rule is most efficient when weights have been optimized with respect to the treatment differences and the correlation between them. 
The Compensatory rule with equal weights (C-E) is most efficient when treatment differences on both outcomes are equally large (treatment differences $3-5$), while the Compensatory rule with unequal weights for uncorrelated outcomes (C-UU) is most efficient under data generating mechanism $8.2$.
The Compensatory rule with unequal weights, optimized for negatively correlated outcomes (C-UC) is most efficient in data generating mechanism $8.1$. 
The Compensatory is less efficient than the Single rule in the scenario with an effect on one outcome only (treatment difference $6$).
Effectively, in this situation the Single rule is the Compensatory rule with optimal weights for this specific scenario $\bm{w}=(1,0)$.
Utilizing the flexibility of the Compensatory rule to tailor weights to anticipated treatment differences and their correlations thus pays off in terms of efficiency.

Note that in practice it may be difficult to accurately estimate treatment differences and correlations in advance.
This uncertainty may result in inaccurate sample size estimates, as demonstrated in Appendix \textit{\nameref{app:compare_designs}}.
The simulations in this appendix also show that the approach can be implemented in designs with interim analyses as well, which is particularly useful under uncertainty about anticipated treatment differences. 
Specifically, we demonstrate that 1) both Type I and Type II error rates increase, while efficiency decreases in a fixed design when the anticipated treatment difference does not correspond to the true treatment difference; and 2) designs with interim analyses could compensate for this uncertainty in terms of error rates and efficiency, albeit at the expense of upward bias.

Further, Appendix \textit{\nameref{app:compare_priors}} shows how prior information influences the properties of decision-making.
Informative priors support efficient decision-making when the prior treatment difference corresponds to the treatment difference in the data.
In contrast, evidence is influenced by dissimilarity between prior hyperparameters and data, and may either increase or decrease 1) the required sample size; and 2) the average posterior treatment effect, depending on the nature of the non-correspondence.

%%%%%%%%%%%%%%%%%%%%%%%%%%%%%%%%%%%%%%%%%%%%%%%%%%%%%%%%%%%%%%%%%%%%%%%%%%%%%
%																			%
%								SECTION 6 									%
%																			%
%%%%%%%%%%%%%%%%%%%%%%%%%%%%%%%%%%%%%%%%%%%%%%%%%%%%%%%%%%%%%%%%%%%%%%%%%%%%%	

\section{Discussion}\label{sec:discussion}

	The current paper presented a Bayesian framework to efficiently combine multiple binary outcomes into a clinically relevant superiority decision. 
	We highlight two characteristics of the approach.

	First, the multivariate Bernoulli model has shown to capture relations properly and support multivariate decision-making. 
	The influence of the correlation between outcomes on the amount of evidence in favor of a specific treatment highlights the urgency to carefully consider these relations in trial design and analysis in practice.

	Second, multivariate analysis facilitates comprehensive decision rules such as the Compensatory rule. 
	More specific criteria for superiority can be defined to ensure clinical relevance, while relaxing conditions that are not strictly needed for clinical relevance lowers the sample size required for error control; A fact that researchers may take advantage of in practice where sample size limitations are common \cite{VandeSchoot2020}.

Several other modeling procedures have been proposed for the multivariate analysis of multiple binary outcomes.
The majority of these alternatives assume a (latent) normally distributed continuous variable. 
When these models rely on large sample approximations for decision-making (such as methods presented by Whitehead et al. \cite{Whitehead2010}, Sozu et al. \cite{Sozu2010,Sozu2016}, and Su et al. \cite{Su2012}; see for an exception Murray et al. \cite{Murray2016}), 
their applicability is limited, since the validity of z-tests for small samples may be inaccurate.
A second class of alternatives uses copula models, which is a flexible approach to model dependencies between multiple univariate marginal distributions.  
The use of copula structures in discrete data can be challenging however \cite{Panagiotelis2012}.
Future research might provide insight in the applicability of copula models for multivariate decision making in clinical trials.

Two additional remarks concerning the number of outcomes should be made.
First, the modeling procedure becomes more complex when the number of outcomes increases, since the number of cells increases exponentially. 
Second, the proposed Compensatory rule has a linear compensatory mechanism.
With two outcomes, the outcomes compensate each other directly and the size of a negative effect is maximized by the size of the positive effect. 
A decision based on more than two outcomes might have the - potentially undesirable - consequence of compensating a single large negative effect by two or more positive effects. 
Researchers are encouraged to carefully think about a suitable superiority definition and might consider additional restrictions to the Compensatory rule, such as a maximum size of individual negative effects.

%\begin{dci}
%	The Authors declare that there is no conflict of interest.
%\end{dci}

%\begin{funding}
%	This work was supported by the Dutch Research Council (NWO) [no. 406.18.505].
%\end{funding}

%\begin{acks}
%	We thank two anonymous reviewers for their helpful comments that greatly improved the presentation of the main ideas in this manuscript.
%\end{acks}

%\begin{sm}
%	The \texttt{R} code used to generate results in Section \textit{\nameref{sec:evaluation}}, Appendix \textit{\nameref{app:compare_designs}}, and Appendix \textit{\nameref{app:compare_priors}} can be found on \url{https://github.com/XynthiaKavelaars/Decision-making-with-multiple-correlated-binary-outcomes-in-clinical-trials}
%\end{sm}

%%%%%%%%%%%%%%%%%%%%%%%%%%%%%%%%%%%%%%%%%%%%%%%%%%%%%%%%%%%%%%%%%%%%%%%%%%%%%
%																			%
%							BIBLIOGRAPHY 									%
%																			%
%%%%%%%%%%%%%%%%%%%%%%%%%%%%%%%%%%%%%%%%%%%%%%%%%%%%%%%%%%%%%%%%%%%%%%%%%%%%%	

\newpage
\bibliographystyle{SageV}
\bibliography{Project1.bib}

\newpage
\begin{appendices}
	
%%%%%%%%%%%%%%%%%%%%%%%%%%%%%%%%%%%%%%%%%%%%%%%%%%%%%%%%%%%%%%%%%%%%%%%%%%%%%
%																			%
%			APPENDIX A Prior specification									%
%																			%
%%%%%%%%%%%%%%%%%%%%%%%%%%%%%%%%%%%%%%%%%%%%%%%%%%%%%%%%%%%%%%%%%%%%%%%%%%%%%
\section{Specification of prior hyperparameters}\label{app:prior}
We might facilitate specification of hyperparameters when we consider the prior distribution of joint success probabilities $\bm{\theta}_{j}$ rather than the prior distribution of joint response probabilities $\bm{\phi}_{j}$.
Here we can utilize the facts that 1) the multivariate beta distribution is a transformation of the Dirichlet distribution; and 2) the parameters of the two distributions are identical \cite{Olkin2015}.

We present the transformation for $K=2$, such that $Q=2^{K}=4$.
The Dirichlet distribution with hyperparameters 
${\bm{\alpha}_{j}}^{0}=(
\alpha_{j,11 \dots 11}^{0}, 
\alpha_{j,11 \dots 10}^{0},
\dots, 
\alpha_{j,00 \dots 01}^{0},
\alpha_{j,00 \dots 00}^{0} 
)$
has the following form:  
	\begin{flalign}\label{eq:k-dir}
	p(\bm{\phi}_{j}|\bm{\alpha}^{0}_{j})=&
	\text{Dirichlet}(\bm{\phi}_{j}|\bm{\alpha}^{0}_{j})&&\\\nonumber
	\propto&
	\phi_{j,1\dots 11}^{\alpha^{0}_{j,1\dots 11}-1} 
	\phi_{j,1\dots 10}^{\alpha^{0}_{j,1\dots 10}-1} 
	\times \dots \times
	\phi_{j,0\dots 01}^{\alpha^{0}_{j,0\dots 01}-1} 
	\phi_{j,0\dots 00}^{\alpha^{0}_{j,0\dots 00}-1}. 
	&&
	\end{flalign}
\noindent Reparametrizing $\bm{\phi}_{j}$ in terms of $\bm{\theta}_{j}$ and integrating $\phi_{j,11}$ out transforms the Dirichlet distribution of posterior $\bm{\phi}_{j}$ to a multivariate beta posterior distribution of success probabilities $\bm{\theta}_{j}$ \cite{Olkin2015}:%
	\begin{flalign}\label{eq:bvbeta}
	p(\bm{\theta}_{j}|\bm{\alpha}_{j}^{n})=&
	\frac{1}{B(\bm{\alpha}_{j}^{n})} \int_\Omega 
	\phi_{j,11}^{\alpha^{n}_{j,11}-1}  \times
	(\theta_{j,1}-\phi_{j,11})^{\alpha^{n}_{j,10}-1} \times
	&\\\nonumber
	& (\theta_{j,2}-\phi_{j,11})^{\alpha^{n}_{j,01}-1} \times  (1-\theta_{j,1}-\theta_{j,2}+\phi_{j,11})^{\alpha^{n}_{j,00}-1}
	\partial\phi_{j,11},&
	\end{flalign}
	\begin{conditions}
		$$\Omega = $$ & $\phi_{j,11}: \text{max}(0,\theta_{j,1}+\theta_{j,2}-1)<\phi_{j,11}<\text{min}(\theta_{j,1},\theta_{j,2})$.\\
	\end{conditions}
\noindent Note that prior $\bm{\theta}_{j}$ also follows a multivariate beta distribution.
However, the multivariate beta distribution cannot be formally used as a prior distribution in posterior computation, since the distribution is marginalized with respect to the information about the relation between success probabilities in $\phi_{j,kl}$.%(i.e. $\phi_{j,11}$ when $K=2$).

\begin{figure}[htbp]
	\centering
	\begin{subfigure}[c]{\linewidth}
		\centering
		\includegraphics[trim=0 50 0 100,clip,width=0.45\linewidth,height=0.24\textheight,keepaspectratio]{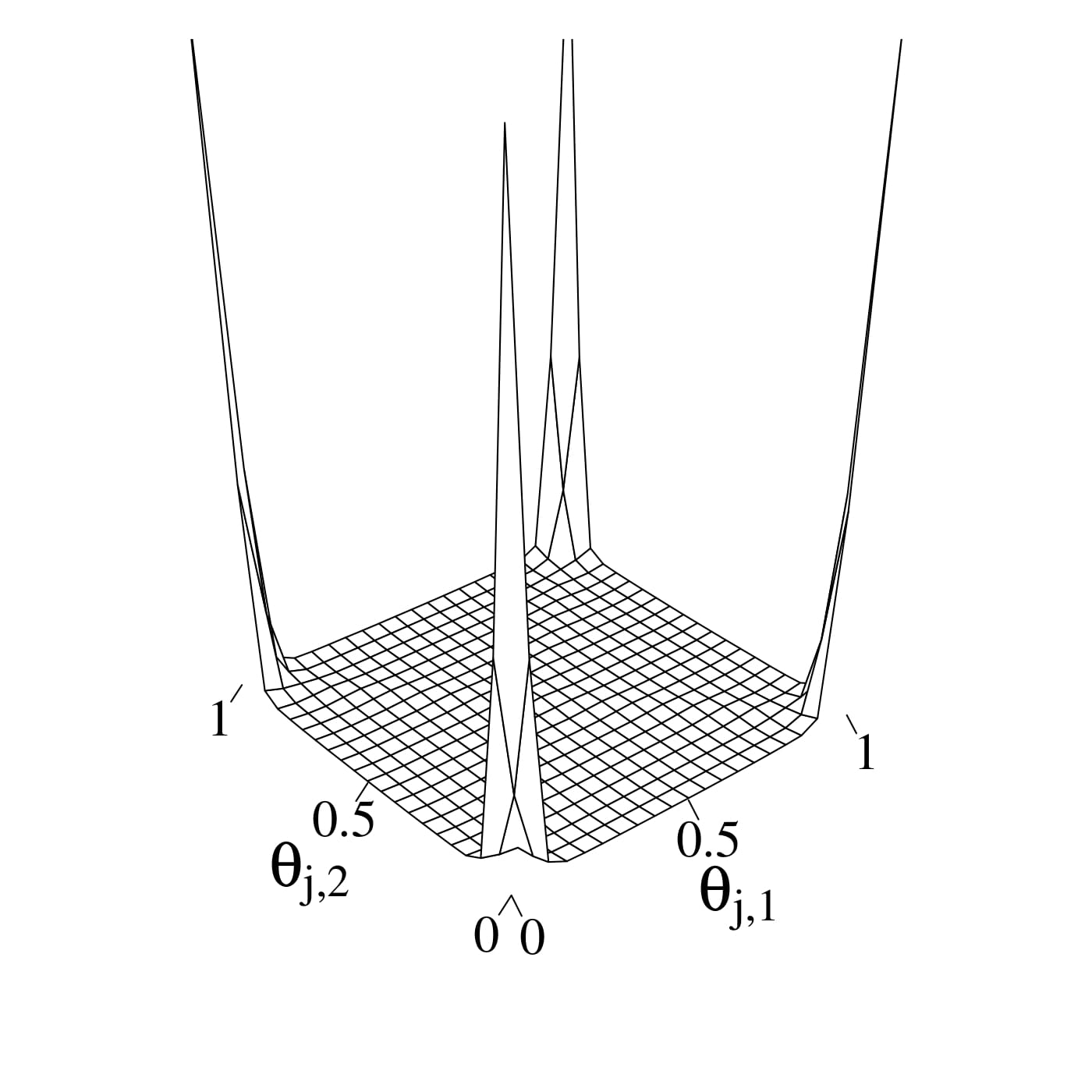}
		\includegraphics[trim=0 50 0 100,clip,width=0.45\linewidth,height=0.24\textheight,keepaspectratio]{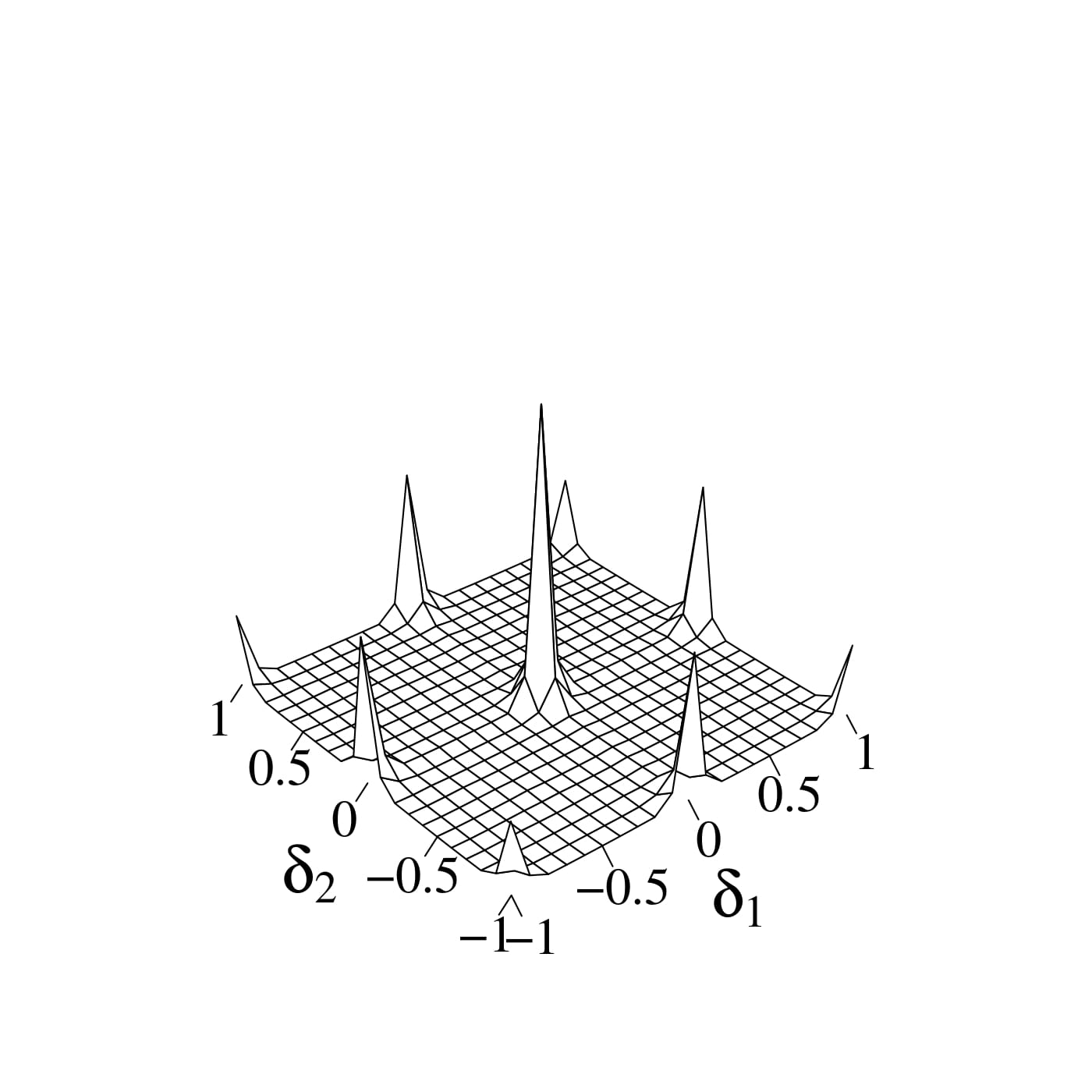}
		\caption{$\bm{\alpha}^{0}_{j}=(0.01,0.01,0.01,0.01)$; $n^{0}_{j}=0.04$}
	\end{subfigure}
	
	\begin{subfigure}[c]{\linewidth}
		\centering
		\includegraphics[trim=0 50 0 100,clip,width=0.45\linewidth,height=0.24\textheight,keepaspectratio]{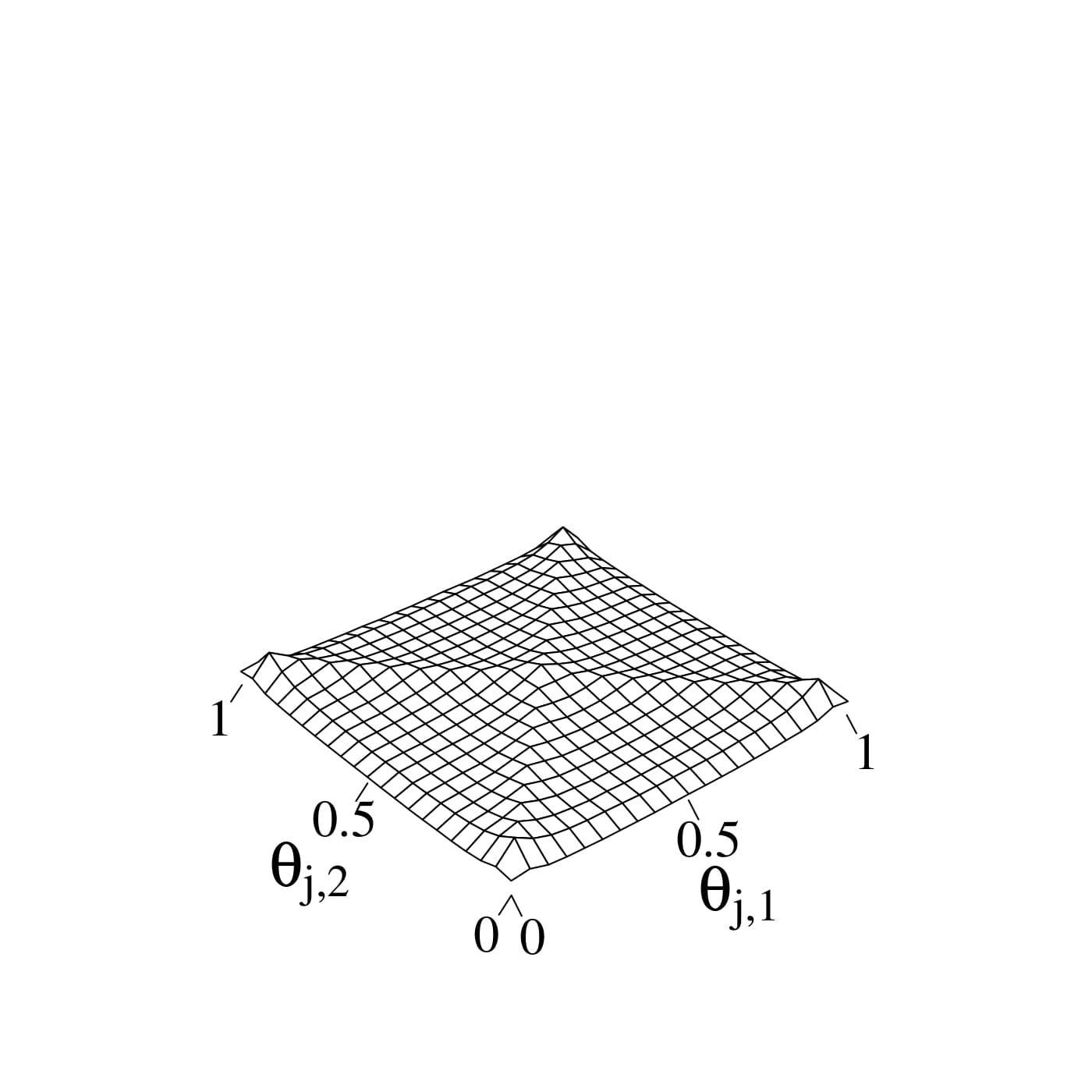}
		\includegraphics[trim=0 50 0 100,clip,width=0.45\linewidth,height=0.24\textheight,keepaspectratio]{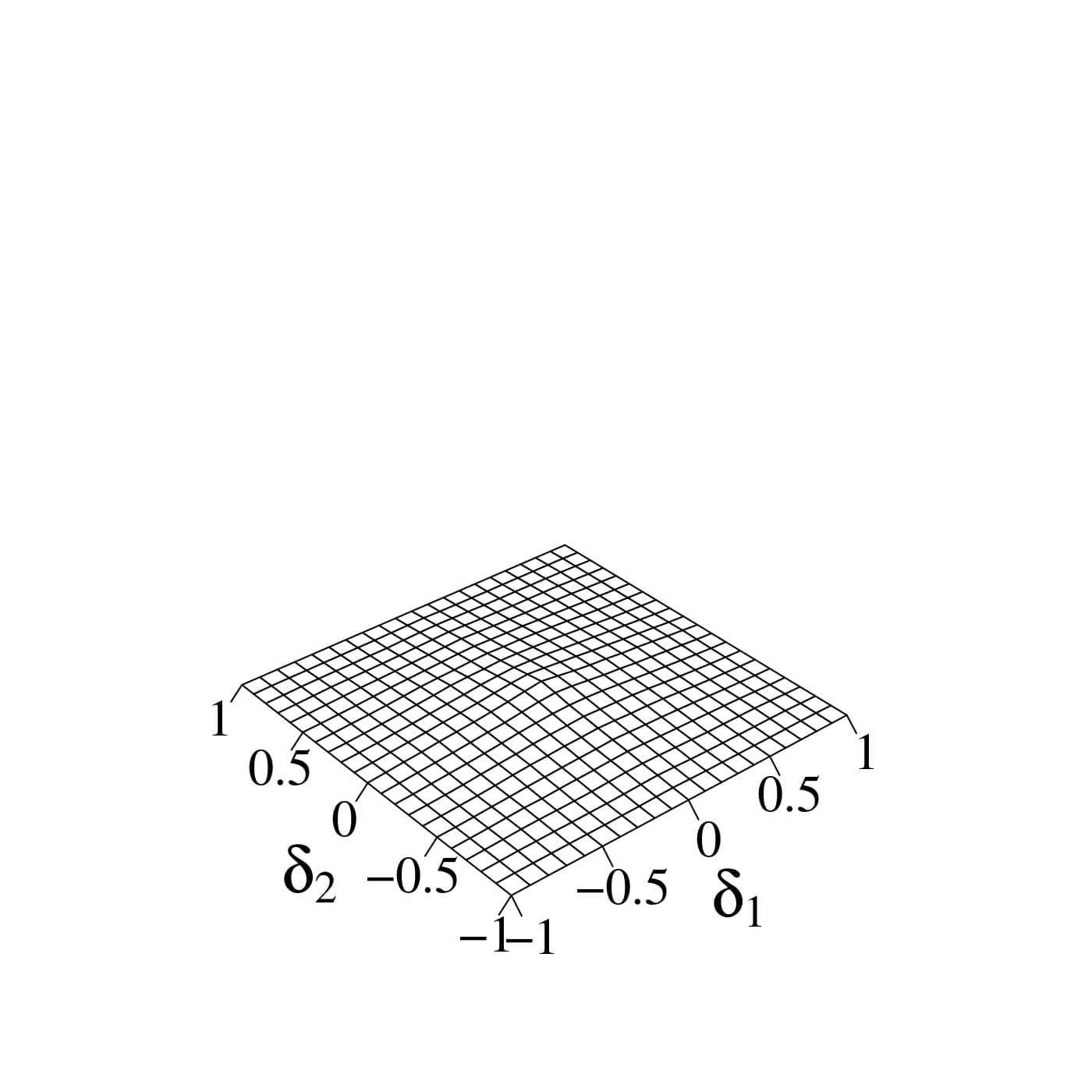}
		\caption{$\bm{\alpha}^{0}_{j}=(0.5,0.5,0.5,0.5)$; $n^{0}_{j}=2$}
		\label{fig:bibeta_b}
	\end{subfigure}
	
	\begin{subfigure}[c]{\linewidth}
		\centering	
		\includegraphics[trim=0 50 0 100,clip,width=0.45\linewidth,height=0.24\textheight,keepaspectratio]{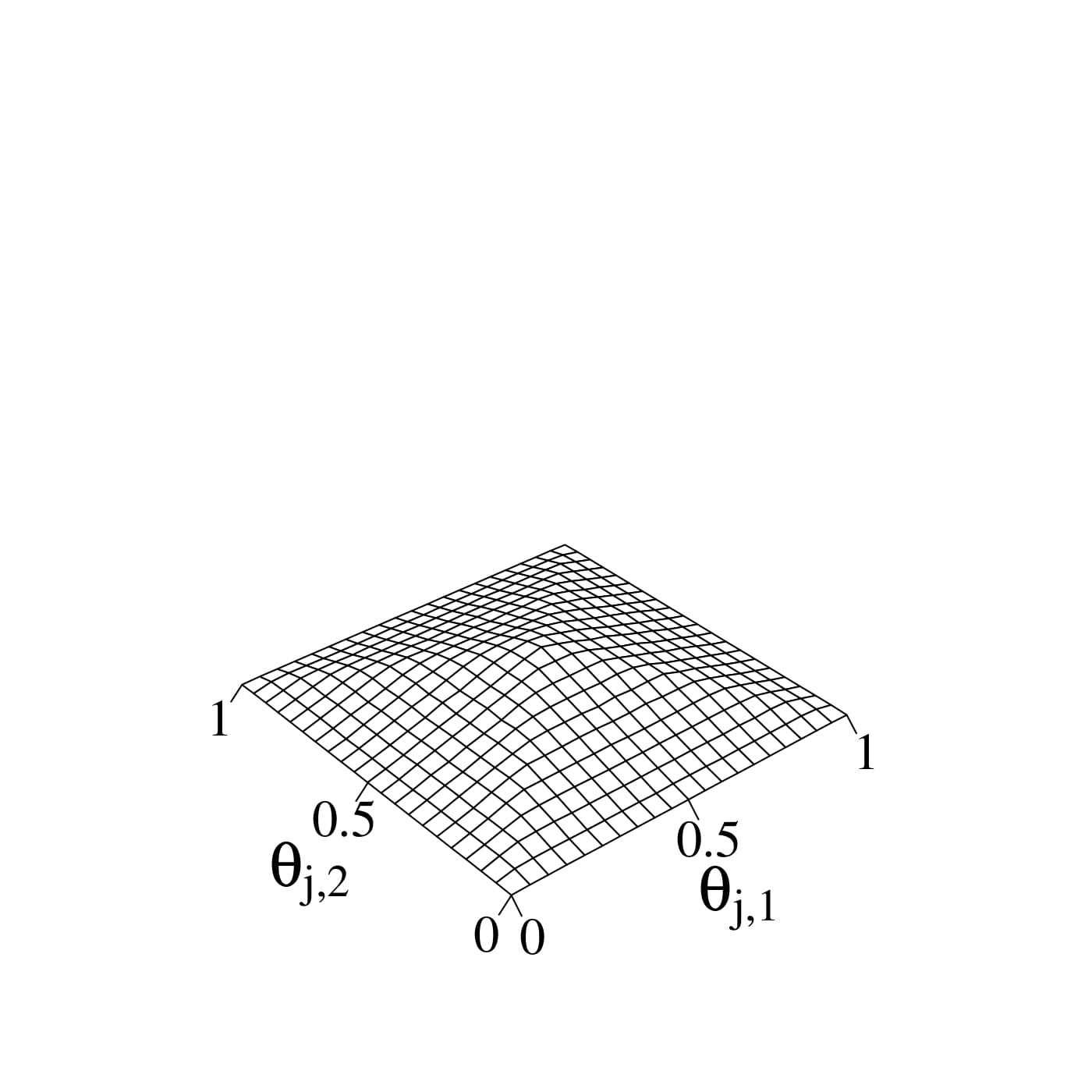}
		\includegraphics[trim=0 50 0 100,clip,width=0.45\linewidth,height=0.24\textheight,keepaspectratio]{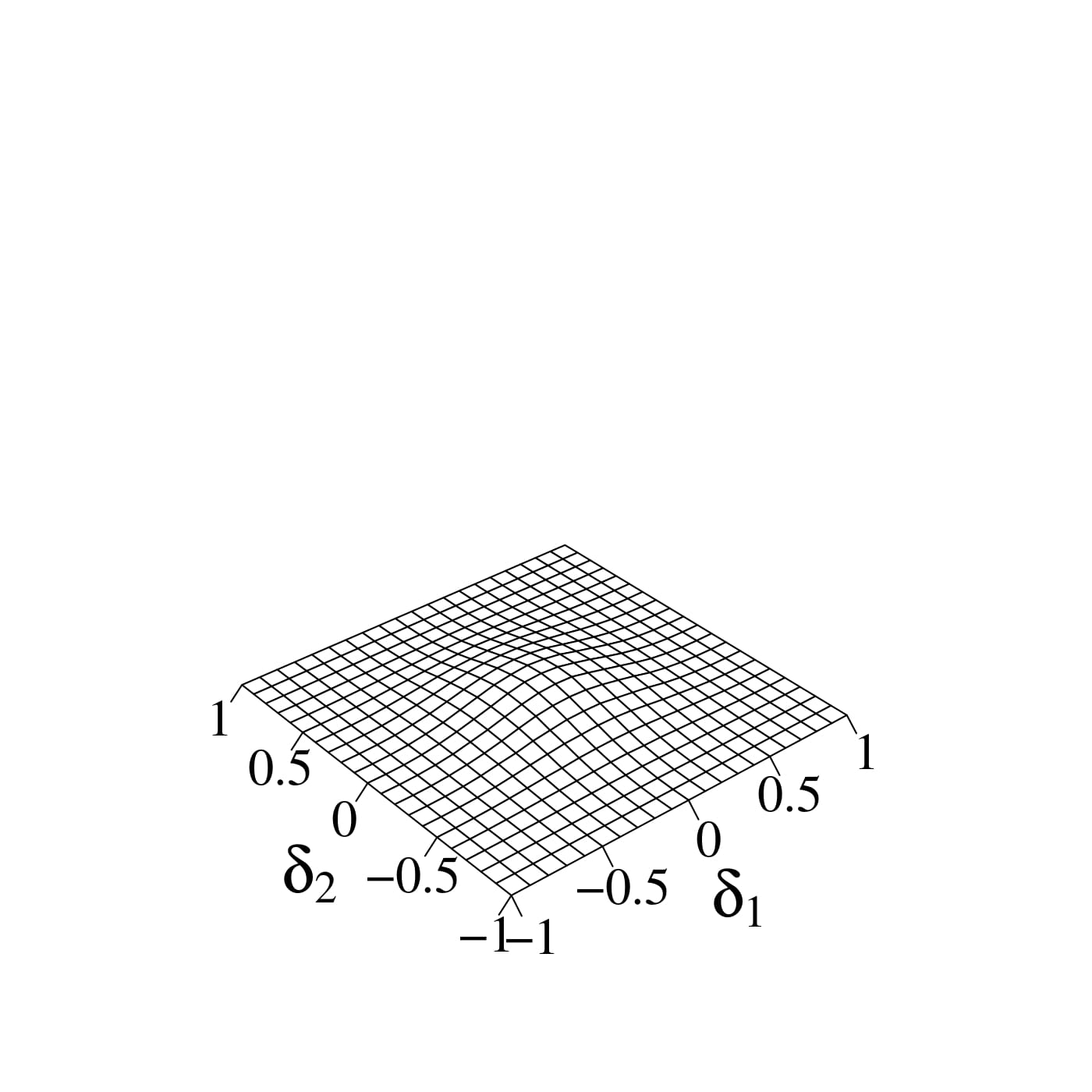}
		\caption{$\bm{\alpha}^{0}_{j}=(1,1,1,1)$; $n^{0}_{j}=4$}
	\end{subfigure}
	
	\begin{subfigure}[c]{\linewidth}
		\centering
		\includegraphics[trim=0 50 0 100,clip,width=0.45\linewidth,height=0.24\textheight,keepaspectratio]{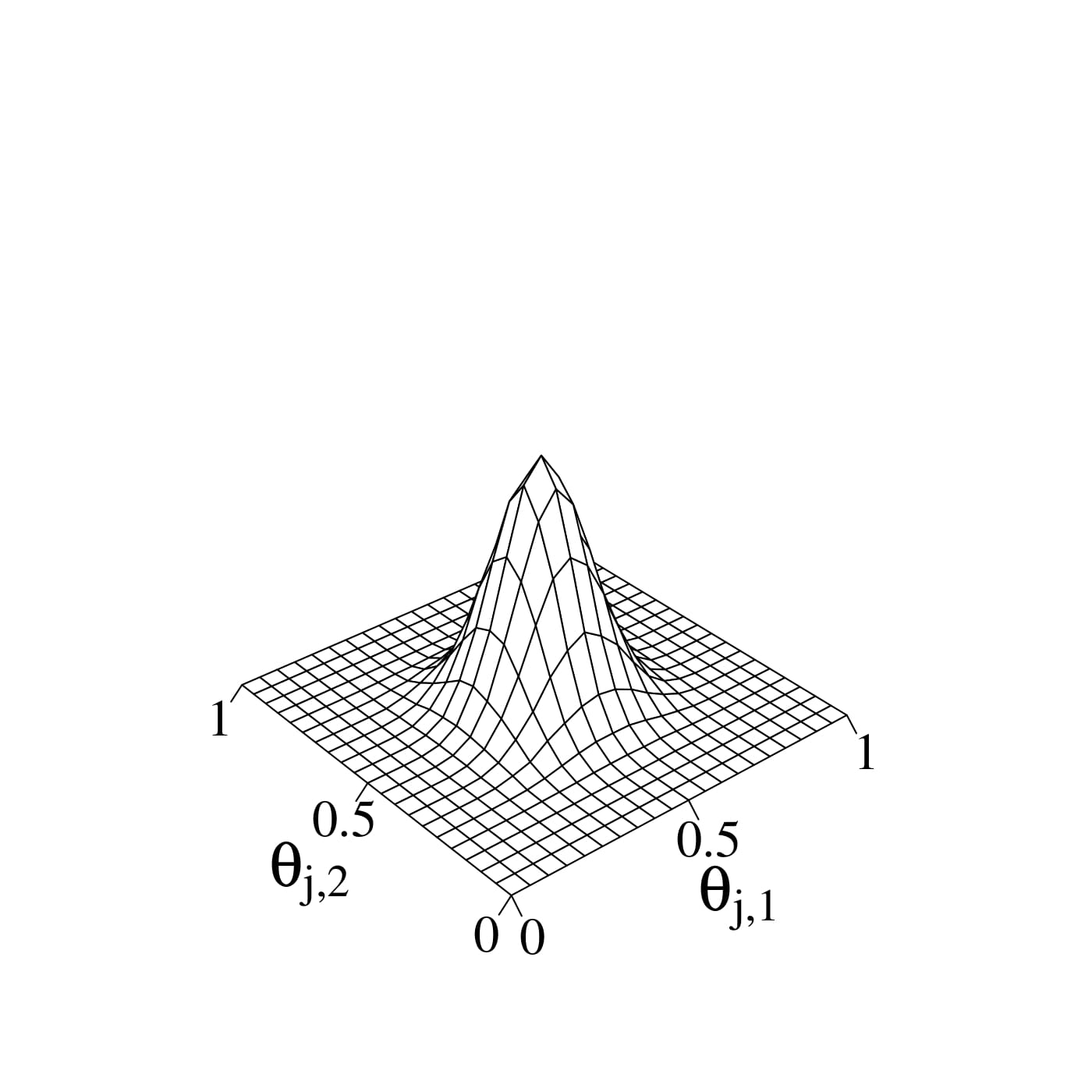}
		\includegraphics[trim=0 50 0 100,clip,width=0.45\linewidth,height=0.24\textheight,keepaspectratio]{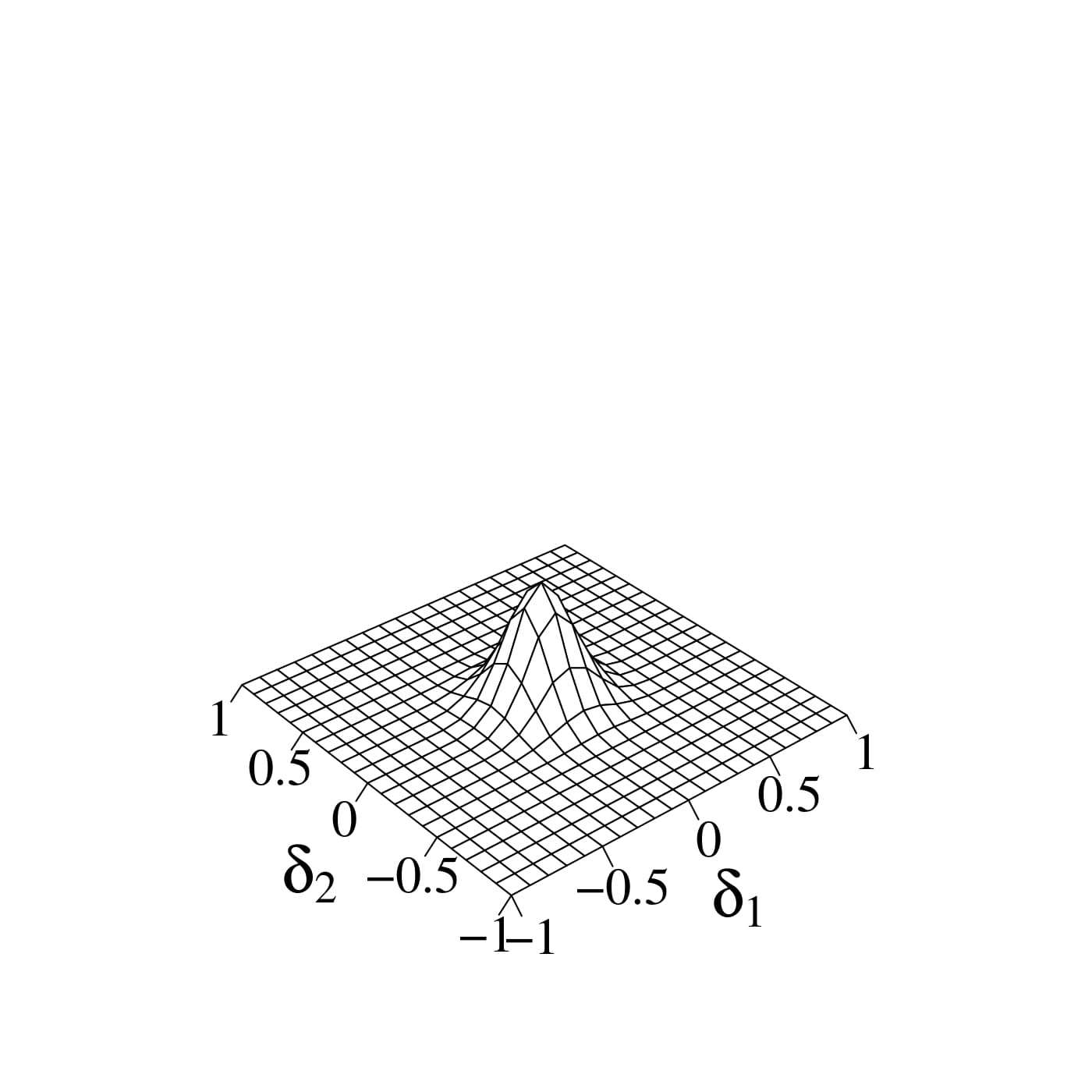}
		\caption{$\bm{\alpha}^{0}_{j}=(5,5,5,5)$; $n^{0}_{j}=20$}
	\end{subfigure}
	
	\caption{Bivariate prior distributions of $\bm{\theta}_{j}$ (left) and $\bm{\delta}=\bm{\theta}_{E}-\bm{\theta}_{C}$ (right) for various specifications of hyperparameters $\bm{\alpha}^{0}_{j}$ and $n^{0}_{j}$ when $K=2$. 
		Prior response probability $\bm{\phi}^{0}_{j}=\frac{1}{4}$.}
	\label{fig:bibeta}
	
\end{figure}

Let us further redefine $\bm{\alpha}^{0}_{j}$ as $n^{0}_{j}\bm{\phi}^{0}_{j}$ to provide an intuitive specification of prior information.
Here $n^{0}_{j}$ reflect the amount of prior information and $\bm{\phi}_{j}^{0}$  reflects the prior means of joint response probabilities $\bm{\phi}_{j}$. 
Prior means $\bm{\phi}_{j}^{0}$ relate directly to the prior means of joint success probabilities $\bm{\theta}^{0}_{j}$ and the prior mean treatment difference $\bm{\delta}^{0}$, since $\theta^{0}_{j,k}$ equals the sum of all elements of $\bm{\phi}^{0}_{j}$ with the $k^{th}$ element of response combination $q$ equal to $1$ and $\delta^{0}_{k}=\theta^{0}_{E,k}-\theta^{0}_{C,k}$.
The following paragraph lists the influence of hyperparameters $n^{0}_{j}\bm{\phi}^{0}_{j}$ on the shape of the prior distributions of success probabilities $\bm{\theta}^{0}_{j}$ and treatment differences $\bm{\delta}$.

\begin{figure}[htbp]
	\centering
	\includegraphics[width=0.5\textwidth,keepaspectratio]{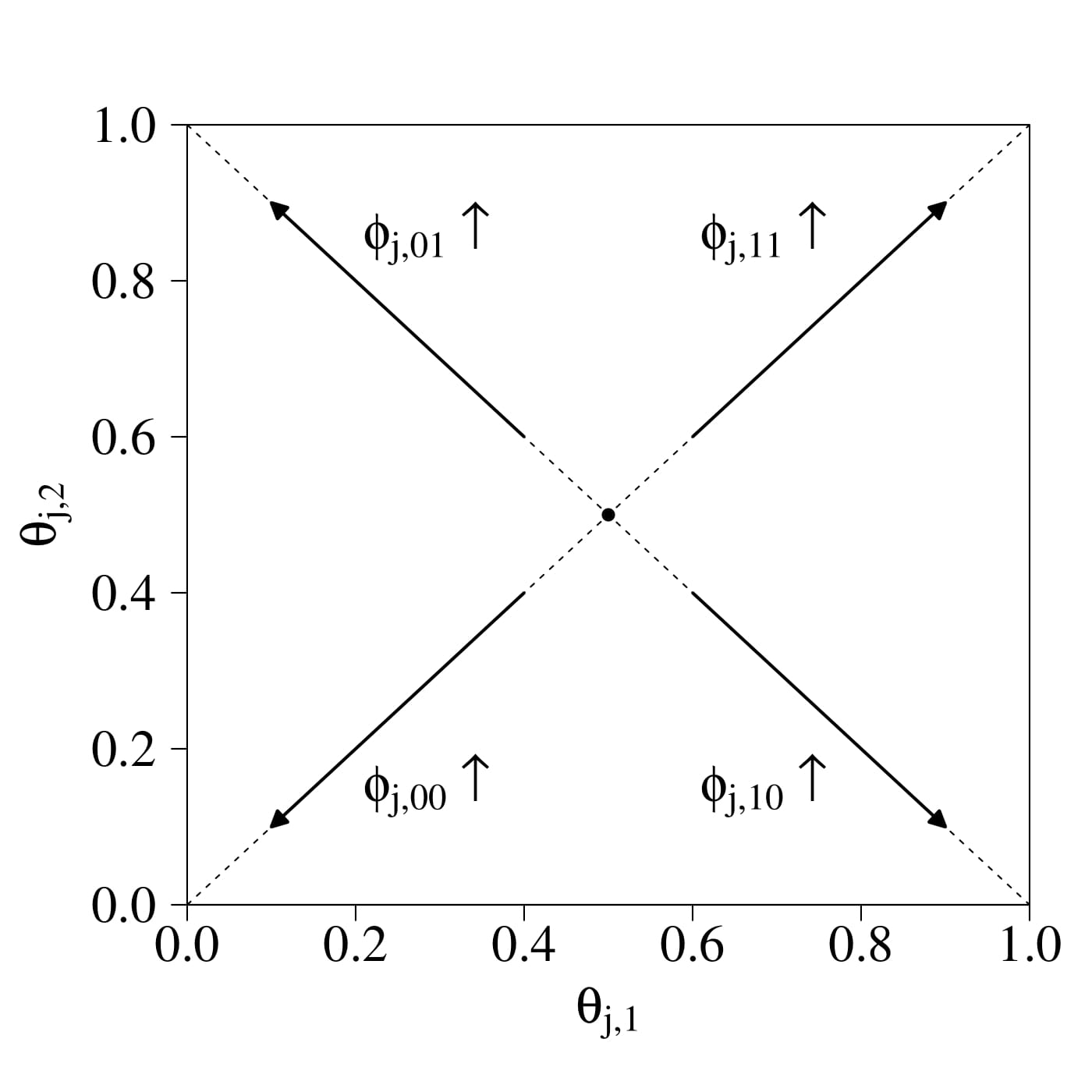}
	\caption{The influence of $\bm{\phi}_{j}$ on the bivariate beta distribution of $\bm{\theta}_{j}$ for two outcomes ($K=2$).}
	\label{fig:mode}
\end{figure}

\begin{enumerate}
	
	\item 
	The amount of prior information in $n^{0}_{j}$ determines the spread of the prior distribution of treatment $j$, as visualized in Figure \ref{fig:bibeta}.  
	Large $n^{0}_{j}$ results in a peaked distribution that reflects more prior information, whereas small $n^{0}_{j}$ results in a distribution with heavy tails that conveys little prior information. 
	Parameter $n^{0}_{j}$ can be considered a prior sample size, where each observation has the same influence on the decision as one joint response $\bm{x}_{j,i}$.
	
	\item 
	Mean prior success probabilities $\bm{\theta}^{0}_{j,k}$ define the center of the prior distribution of success probabilities $\bm{\theta}_{j}$, as visualized in Figure \ref{fig:mode}. 
	Similarly, mean prior treatment differences $\bm{\delta}^{0}_{k}$ reflect the center of the prior distribution of $\bm{\delta}$.
	When $n^{0}_{E}=n^{0}_{C}$ and $\bm{\phi}^{0}_{E}=\bm{\phi}^{0}_{C}$, the prior distribution of the treatment difference $\bm{\delta}$ is centered around the origin (i.e. $\bm{\delta}^{0}=\bm{0}$). 
	
	\item 
	The size of $\phi^{0}_{j,kl}$ relative to $\theta^{0}_{j,k}\theta^{0}_{j,l}$ determines the prior correlation between $\theta_{j,k}$ and $\theta_{j,l}$ \cite{Olkin2015}:%
		\begin{flalign}\label{eq:rho_bibeta}
		\rho_{\theta_{j,k}\theta_{j,l}}=&
		\frac{\phi^{0}_{j,kl}-\theta^{0}_{j,k}\theta^{0}_{j,l}}
		{\sqrt{\theta^{0}_{j,k}(1-\theta^{0}_{j,k})\theta^{0}_{j,l}(1-\theta^{0}_{j,l})}}.&&
		\end{flalign}
	As follows from Equation \ref{eq:rho_bibeta}, $\theta_{j,k}$ and $\theta_{j,l}$ are independent a priori if $\phi^{0}_{j,kl}=\theta^{0}_{j,k}\theta^{0}_{j,l}$. 
	When $\phi^{0}_{j,kl}>\theta^{0}_{j,k}\theta^{0}_{j,l}$, parameters are positively correlated, while parameters are negatively correlated when $\phi^{0}_{j,kl}<\theta^{0}_{j,k}\theta^{0}_{j,l}$.
	
\end{enumerate}

When prior information is known, the introduced properties can be used to make informative choices regarding these parameters.
In absence of prior information however, a reasonable prior distribution of success probabilities $\bm{\theta}_{j}$ has 1) a small $n^{0}_{j}$, such that the impact on the decision is limited; and 2) $\theta_{j,k}=\frac{1}{2}$, such that successes and failures are equally likely a priori for all treatments $j$ and all outcomes $k$. 
Although mathematically straightforward, we remark that estimating prior hyperparameters in practice can be challenging when $K$ is large.

%%%%%%%%%%%%%%%%%%%%%%%%%%%%%%%%%%%%%%%%%%%%%%%%%%%%%%%%%%%%%%%%%%%%%%%%%%%%%
%																			%
%			APPENDIX B Specification of efficiency weights					%
%																			%
%%%%%%%%%%%%%%%%%%%%%%%%%%%%%%%%%%%%%%%%%%%%%%%%%%%%%%%%%%%%%%%%%%%%%%%%%%%%%
%\newpage
\section{Specification of efficiency weights}\label{app:weights}
	Finding weights that maximize efficiency requires maximizing the following function with respect to $\bm{w}$: 
	\begin{flalign}\label{eq:max_weights}
	\sum_{\bm{s}^{*}_{E},\bm{s}^{*}_{C}}&
	P(\bm{\delta}\in \mathcal{S}_{Compensatory}(\bm{w})|\bm{s}^{*}_{E},\bm{s}^{*}_{C}) \times
	P(\bm{s}^{*}_{E},\bm{s}^{*}_{C}|\bm{\theta}^{T}_{E},\bm{\theta}^{T}_{C},\bm{\rho}^{T}_{\delta_{k},\delta_{l}})&&\\\nonumber
	=&  P(\bm{\delta}\in \mathcal{S}_{Compensatory}(\bm{w})|\bm{\theta}^{T}_{E}, \bm{\theta}^{T}_{C}, \bm{\rho}^{T}_{\delta_{k},\delta_{l}})&&
	\end{flalign}
	\begin{conditions}
		$$\bm{s}^{*}_{E}$$ 
		& and $\bm{s}^{*}_{C}$ 
		are the anticipated response frequencies before data collection\\
		$$\bm{\theta}^{T}_{E}$$ & and $\bm{\theta}^{T}_{C}$
		are the treatment effects in the population\\
		$$\bm{\rho}^{T}_{\delta_{k},\delta_{l}}$$ & is the correlation between $\delta_{k}$ and $\delta_{l}$ in the population. \\			
	\end{conditions}
No analytical solution for Equation \ref{eq:max_weights} exists.
We can however obtain a solution for $\bm{w}$ using large sample theory, which dictates that the posterior distribution of $\bm\delta$ can be approximated with a multivariate normal distribution in case of a sufficiently large sample:
	\begin{flalign}
	\bm{\delta} \sim MVN(\bm{\mu}, \bm{\Sigma})
	\end{flalign}
	\begin{conditions}
		\bm{\mu}= & $(\mu_{1},\dots,\mu_{K})$ and \\
		\bm{\Sigma} & has diagonal elements $\bm{\sigma}^{2}=(\sigma^{2}_{1},\dots,\sigma^{2}_{K})$ and off-diagonal elements $\bm{\sigma}_{kl}$.\\
	\end{conditions}
\noindent Consequently, the linear combination $\sum_{k=1}^{K} w_k\delta_k$ has an approximate normal posterior distribution with mean $\sum_{k=1}^{K} w_{k}\mu_{k}$ and variance $\sum_{k=1}^{K} w^{2}_{k}\sigma^{2}_{k}+2\mathop{\sum\sum}\limits_{k < l} w_{k} w_{l} \sigma_{kl}$.
The probability that $\sum_{k=1}^{K} w_{k}\delta_{k}>0$ then follows from the cumulative normal distribution:
	\begin{flalign}\label{eq:cumnorm}
	P(\sum^{K}_{k=1} w_{k} \delta_{k} > 0) =
	1-\Phi \left( \frac{0-\sum_{k=1}^{K} w_{k}\mu_{k}}
	{\sqrt{\sum_{k=1}^{K} w^{2}_{k}\sigma^{2}_{k}+2\mathop{\sum\sum}\limits_{k < l} w_{k} w_{l} \sigma_{kl}}} \right).&&
	\end{flalign}
Weights $\bm{w}$ that maximize the probability in Equation \ref{eq:cumnorm} result in maximal efficiency. 
In practice, computing efficient weights is less straightforward since $\bm{\mu}$ and $\bm{\Sigma}$ are unknown.
To facilitate the choice of these parameters for the construction of a normal posterior distribution of $\bm{\delta}$, we may consider hypothetical datasets of expected joint response frequencies $\bm{s}^{*}_{j}$ for both treatments $j$.
These frequencies can be used to obtain a sample of $\bm{\delta}$, that is assumed to follow a normal distribution when the sample size $n_{j}$ and the number of draws are sufficiently large. 
Such a sample provides estimates of $\bm{\mu}$, $\bm{\sigma}^{2}$ and $\bm{\sigma}_{kl}$, that can be plugged in in Equation \ref{eq:cumnorm}.
We provide an example data configuration for $K=2$ in Table \ref{tab:data}.
This hypothetical dataset would result in $\bm{\mu}=(0.24,0.08)$, $\bm{\sigma}^{2}=(0.005,0.005)$ and $\sigma_{12}=-0.001$, such that optimal weights equal $\bm{w}=(0.64,0.36)$.

\begin{table}[htbp]
	\small\sf\centering
	\caption{Example configuration of anticipated joint response frequencies $\bm{s}^{*}_{E}$ and $\bm{s}^{*}_{C}$ for approximation of $\bm{\mu}$,$\bm{\sigma}^{2}$, and $\sigma_{12}$ for two outcomes.}
	\label{tab:data}
	
	\begin{tabular}{crr p{0.05\linewidth} crr}
		
		\toprule
		& $s^{*}_{E,1}=1$ & $s^{*}_{E,1}=0$ & & & $s^{*}_{C,1}=1$ & $s^{*}_{C,1}=0$\\ 
		
		\midrule
		$s^{*}_{E,2}=1$ & 262 & 278 & & $s^{*}_{C,2}=1$ & 102 & 358 \\
		$s^{*}_{E,2}=0$ & 358 & 102 & & $s^{*}_{C,2}=0$ & 278 & 262 \\
		\bottomrule
	\end{tabular}
	
\end{table}

The procedure to find efficient weights simplifies when treatment differences are uncorrelated, i.e. when $\sigma_{kl}=0$.
Maximum evidence is then obtained when weights $\bm{w}$ are proportional to treatment difference $\bm{\delta}$ and $\sigma^{2}_{k}=\sigma^{2}_{l}$.
For example, when $\bm{\delta}=(0.30,0.10)$, weights $\bm{w}=(0.75,0.25)$ are optimal.

%%%%%%%%%%%%%%%%%%%%%%%%%%%%%%%%%%%%%%%%%%%%%%%%%%%%%%%%%%%%%%%%%%%%%%%%%%%%%
%																			%
%			APPENDIX D Implementation of the framework in group-sequential  %
%			and adaptive designs											%
%																			%
%%%%%%%%%%%%%%%%%%%%%%%%%%%%%%%%%%%%%%%%%%%%%%%%%%%%%%%%%%%%%%%%%%%%%%%%%%%%%
\section{Implementation of the framework in group sequential and adaptive designs}\label{app:implementation_adaptive}

The current appendix presents an algorithm with the procedure to arrive at a decision using the multivariate analysis procedure for a group sequential or adaptive design. 

\begin{algorithm}
	\begin{tabular}{p{\textwidth}}
		\caption{Decision procedure for a group sequential or adaptive design}
		\label{alg:interim}
		
		\begin{enumerate}[label={\arabic*},leftmargin=7pt,topsep=0pt,parsep=0pt,labelindent=0pt,itemindent=0pt,listparindent=-5pt]
			\item 
			\underline{\textbf{Initialize}}		
			
			\begin{enumerate}[label={},leftmargin=7pt,topsep=0pt,parsep=0pt,labelindent=0pt,itemindent=0pt,listparindent=-5pt]
				\item 
				\begin{enumerate}[label={\alph*},leftmargin=7pt,topsep=0pt,parsep=0pt,labelindent=0pt,itemindent=0pt,listparindent=-5pt]
					\item 
					Choose decision rule
					
					\begin{enumerate}[label={}]
						\item 
						\textbf{if} Compensatory
						\textbf{then} specify weights $\bm{w}$	
						
						\item 
						\textbf{if} Single
						\textbf{then} specify $k$	
						
						\item 
						\textbf{end if}
						
					\end{enumerate}
					
				\end{enumerate}
				
				\item
				\textbf{for} each treatment $j \in	\{E,C\}$
				\textbf{do}
				
				\begin{enumerate}[label={\alph*},leftmargin=7pt,topsep=0pt,parsep=0pt,labelindent=0pt,itemindent=0pt,listparindent=-5pt]
					\setcounter{enumiii}{1}
					\item 
					Choose prior hyperparameters $\bm{\alpha}^{0}_{j}$ %(see Appendix \textit{\nameref{app:prior}})
					
				\end{enumerate}
				\item 
				\textbf{end for}
				
				\begin{enumerate}[label={\alph*},leftmargin=7pt,topsep=0pt,parsep=0pt,labelindent=0pt,itemindent=0pt,listparindent=-5pt]
					\setcounter{enumiii}{2}
					\item 
					Choose Type I error rate $\alpha$ and power $1-\beta$ 
					
					\item 
					Choose number of interim analyses $M$
					
					\item 
					Determine decision threshold $p_{cut}$
					
					\item
					Determine vector of sample sizes $\bm{n}^{(.)}_{j}$ of length $M$
					
					\begin{enumerate}[label={},leftmargin=7pt,topsep=0pt,parsep=0pt,labelindent=0pt,itemindent=0pt,listparindent=-5pt]
						\item 
						\textbf{if} group sequential design
						\textbf{then}
						$\bm{n}^{SD}_{j} = n^{FD}_{j} \times \bm{n}_{ratio}$,
						where $n^{FD}_{j}$ reflects the required sample size for a fixed design and $\bm{n}_{ratio}$ reflects $M$ proportions of the final sample size at which to perform interim analyses 	
						
						\item 
						\textbf{else if} adaptive design
						\textbf{then} define $\bm{n}^{AD}_{j}$ according to desired monitoring scheme
						
						\item
						\textbf{end if}
						
					\end{enumerate}
					
				\end{enumerate}
			\end{enumerate}
			\item 
			\underline{\textbf{Perform interim analyses}}
			
			\begin{enumerate}[label={},leftmargin=7pt,topsep=0pt,parsep=0pt,labelindent=0pt,itemindent=0pt,listparindent=-5pt]
				\item 
				$m \leftarrow 0$
				
				\item
				\textbf{repeat}
				
				\begin{enumerate}[label={\alph*},leftmargin=7pt,topsep=0pt,parsep=0pt,labelindent=0pt,itemindent=0pt,listparindent=-5pt]
					\item $m \leftarrow m + 1$
					
					\item $n_{j} \leftarrow m^{th}$ element of $\bm{n}_{j}^{(.)}$
					
					\item Collect data and evaluate evidence via Step 2 of Algorithm \ref{alg:fixed}
					
				\end{enumerate}
				\item 
				\textbf{until}
				$m = M$ 
				\textbf{or} 
				$P(\bm{\delta} \in \mathcal{S}_{Sup}|\bm{s}_{E}, \bm{s}_{C}) > p_{cut}$
			\end{enumerate}
			
			\item 
			\underline{\textbf{Make final decision}}
			
			\begin{enumerate}[label={},leftmargin=7pt,topsep=0pt,parsep=0pt,labelindent=0pt,itemindent=0pt,listparindent=-5pt]
				\item 
				\textbf{if} $P(\bm{\delta} \in \mathcal{S}_{Sup}|\bm{s}_{E}, \bm{s}_{C}) > p_{cut}$
				\textbf{then} conclude superiority
				
				\item 
				\textbf{else} conclude non-superiority
				
				\item 
				\textbf{end if}
			\end{enumerate}
		\end{enumerate}\\
	\end{tabular}
\end{algorithm}

%%%%%%%%%%%%%%%%%%%%%%%%%%%%%%%%%%%%%%%%%%%%%%%%%%%%%%%%%%%%%%%%%%%%%%%%%%%%%
%																			%
%			APPENDIX D Comparison of trial designs  						%
%																			%
%%%%%%%%%%%%%%%%%%%%%%%%%%%%%%%%%%%%%%%%%%%%%%%%%%%%%%%%%%%%%%%%%%%%%%%%%%%%%

\section{Numerical evaluation: Comparison of trial designs}\label{app:compare_designs}

The Section \textit{\nameref{sec:evaluation}} showed how accurate decision error rates could be obtained with the proposed framework under a fixed design. 
However, the realization of adequate error rates and efficient decisions depends on the accuracy of sample sizes, and hence on adequate estimates of anticipated treatment differences and correlations. 
A design based on interim analyses might improve statistical inference under parameter uncertainty, which is especially relevant for the estimation of multiple parameters in multivariate analysis \cite{Jennison1999,Berry2010}.
Such trials monitor incoming data and terminate data collection as soon as evidence exceeds a prespecified decision threshold. 
In the current paper, we make a sharp distinction between two of these designs types: adaptive and group sequential designs. 
Here, adaptive designs evaluate the data according to an interim monitoring scheme that does not rely on parameter estimates. 
Such a monitoring scheme and a decision threshold suffice to start data collection.
These designs allow for both early and late termination if the treatment effect appears larger or smaller than anticipated respectively. 
On the downside, efficiency may be compromised if the number of interim analyses high: A (very) strict decision threshold is then needed to the control Type I error rate under repeated decision-making \cite{Sanborn2014,Rouder2014,Shi2019}.

In contrast to adaptive designs, group sequential designs rely on anticipated parameters to estimate a maximum sample size in advance, which is often similar to the sample size of a fixed design. 
Interim analyses are performed at (a limited number of) prespecified proportions of this sample size to allow for early termination.
The potential for late termination is limited with such a setup \cite{Jennison1999}.
To control Type I error rates adequately, decision thresholds $p_{cut}$ should be adjusted to the number of interim analyses, and may differ per interim analysis \cite{Jennison1999,Shi2019}. 
In practice the distinction between group sequential and adaptive designs is less sharp than presented here: Expectations about parameters often (roughly) inform the monitoring scheme in adaptive stopping to limit the number of interim analyses, while conservative parameter estimates allow for late termination of group sequential trials.

In the current appendix, we demonstrate how 1) error rates are influenced by uncertainty about parameters in a priori sample size estimation; and 2) designs with and without interim analyses perform under this uncertainty.
We considered seven different designs:

\begin{enumerate}
	\item A fixed design with sample size $n^{FD}_{j}$ computed with three different anticipated treatment differences $\bm{\delta}^{n}$: %
	\begin{enumerate}%
		\item True treatment differences ($\bm{\delta}^{n} = \bm{\delta}^{T}$)
		\item Overestimated treatment differences ($\bm{\delta}^{n} = \bm{\delta}^{T} + (0.10,0.10)$)
		\item Underestimated treatment differences ($\bm{\delta}^{n} = \bm{\delta}^{T} - (0.10,0.10)$)
	\end{enumerate}%
	\item A group sequential design with a maximum of $M=3$ analyses, evaluated at sample sizes $\bm{n}^{SD}_{j}$ or until superiority is concluded. These sample sizes are computed with three different anticipated treatment differences $\bm{\delta}^{n}$: %
	\begin{enumerate}%
		\item True treatment differences ($\bm{\delta}^{n} = \bm{\delta}^{T}$)
		\item Overestimated treatment differences ($\bm{\delta}^{n} = \bm{\delta}^{T} + (0.10,0.10)$)
		\item Underestimated treatment differences ($\bm{\delta}^{n} = \bm{\delta}^{T} - (0.10,0.10)$)
	\end{enumerate}%
	\item An adaptive design with a maximum of $M=136$ analyses, evaluated at sample sizes $n^{AD}_{j}=n_{j,1},\dots,n_{j,M}$, or until superiority is concluded.
	The first interim analysis is performed at $n^{AD}_{j,1}=5$ and monitors every observation until $50$ observations have been made. 
	Then the interim group size increases to $5$ until $n^{AD}_{j,M}=500$, such that $\bm{n}^{AD}_{j}=(5,6,\dots,49,50,55,\dots,500)$. 
	
\end{enumerate}

\paragraph{Group sequential design}
We set up a group sequential design with equally spaced interim analyses using the \texttt{gsDesign} package \cite{Anderson2016}.
We based computations of interim sample sizes $\bm{n}^{SD}_{j}$ and interim decision threshold $p_{cut}^{SD}$ on the sample size of a fixed design, $n^{FD}_{j}$, using the default settings of the \texttt{gsDesign()} function for a one-sided test with $\alpha=.05$ and $\beta=.20$.

\paragraph{Adaptive design}
Decision threshold $p_{cut}^{AD}$ was calibrated to reflect the desired Type I error rate $\alpha$. 
The procedure involves repeatedly evaluating a large number of simulated samples from the distributions of least favorable values at sample sizes $\bm{n}^{AD}_{j}$ at different values of $p_{cut}$, and selecting the decision threshold for which the empirical Type I error rate corresponds to $\alpha$.

\paragraph{Data generation and evaluation}
We generated $5,000$ samples to compare the seven trial designs for the Compensatory decision rule with equal weights (Comp-E; $\bm{w}=(0.50,0.50)$) and an uninformative prior distribution ($\bm{\alpha}^{0}_{j}=(0.01,0.01,0.01,0.01)$).
We used decision thresholds $p_{cut}^{FD}=0.95$,  $p_{cut}^{SD}=0.98$, and $p_{cut}^{AD}=0.9968$, 
The generated datasets were evaluated using the procedure in Algorithm \ref{alg:interim}.

\subsection{Results}
Tables \ref{tab:CompareDesigns_pSup}, \ref{tab:CompareDesigns_nStop}, and \ref{tab:CompareDesigns_bias} present the results of the comparison of designs.
The performance of fixed and group sequential designs depends on the correspondence between parameter estimates for sample size estimation and the true parameters. 
When parameters were specified correctly (i.e. $\bm{\delta}^{n}=\bm{\delta}^{T}$), both designs resulted in a satisfactory Type I error rate and power (Table \ref{tab:CompareDesigns_pSup}). 
The group sequential design was generally more efficient than the fixed design (Table \ref{tab:CompareDesigns_nStop}).
In this situation, the adaptive design was less efficient than the other designs, since 1) trials are free to continue until superiority has been concluded, resulting in a high (but uncontrolled) power at the expense of a larger sample size; and 2) the increased number of interim analyses in an adaptive design requires a higher decision threshold, which accompanies - on average - a larger average sample size to conclude superiority.

The benefits of interim analyses in terms of decision error rates and efficiency were particularly apparent when anticipated treatment differences did not correspond to the true treatment difference.
A group sequential design is mainly advantageous over a fixed design when sample sizes were based on underestimated treatment differences (i.e. $\bm{\delta}^{n}<\bm{\delta}^{T}$): Probabilities to conclude superiority correctly were well above the planned $.80$ in both designs, but the group sequential design is more efficient.
The adaptive design was especially powerful when anticipated treatment differences were overestimated (i.e. $\bm{\delta}^{n}>\bm{\delta}^{T}$): Both the fixed and group sequential design had a limited power to conclude superiority.

While the adaptive and group sequential designs outperformed the fixed design in terms of power and efficiency under parameter uncertainty, they do result in upward bias (Table \ref{tab:CompareDesigns_bias}).
This effect can be attributed to two different aspects of these designs.
First, the effect is most apparent when the sample size is underestimated ($\bm{\delta}^{n} > \bm{\delta}^{T}$)  . 
Here, the studies that conclude superiority are those with early stops, \textit{because} their effect size at the termination point is larger than the true effect size \cite{Emerson2007}.
The effect size is then averaged over a selection of the (upper part of the) sampling distribution of treatment differences. 
When sample sizes are sufficiently large to allow for timely ($\bm{\delta}^{n} = \bm{\delta}^{T}$) or late ($\bm{\delta}^{n} < \bm{\delta}^{T}$) terminations, these high effect sizes are partially compensated by the samples with smaller effect sizes \cite{Schoenbrodt2017}.

Second, when naively pooled, average effect sizes from trials that stopped early for efficacy are affected by instability of treatment effects early in data collection \cite{Goodman2007,Senn2014,Zhang2012,Schou2013}. 
With few data points, new observations are quite influential resulting in large variation around the treatment effect in the sample. 
In contrast, the treatment effect estimate stabilizes as data accumulate \cite{Zhang2012}.
Trials that allow for early efficacy stopping exploit the variation of small samples to stop trials at extreme values, but only from the upper tail of the distribution.
Extreme treatment effects in the other direction - indicating treatment futility - are given the opportunity to regress to the mean by adding new observations. 
Pooling these extreme values from early stops with the stabilized values from later stops then results in an overestimated treatment effect.
Since our adaptive design has more interim analysis in the beginning of data collection, the adaptive design has more opportunities to include extreme values, resulting in a larger bias compared to the group sequential design.

% latex table generated in R 3.5.3 by xtable 1.8-3 package
% Mon Feb 03 07:06:22 2020
\begin{table}[htbp]
	\small\sf\centering
	\caption{P(Conclude superiority) for different trial designs (AD = adaptive design, FD = fixed design, SD = group sequential design) and anticipated treatment differences ($\bm{\delta}^{n}$) after applying the Compensatory decision rule with equal weights.} 
	\label{tab:CompareDesigns_pSup}
	\resizebox{\textwidth}{!}{
		\begin{tabular}{lrp{0.01\textwidth}rrrp{0.01\textwidth}rrr}
		\toprule
		DGM & 
		\multicolumn{2}{l}{AD} &
		\multicolumn{4}{l}{FD} &
		\multicolumn{3}{l}{SD}\\
		& & & 
		\multicolumn{1}{l}{$\bm{\delta}^{n} = \bm{\delta}^{T}$} & 
		\multicolumn{1}{l}{$\bm{\delta}^{n} < \bm{\delta}^{T}$} &
		\multicolumn{1}{l}{$\bm{\delta}^{n} > \bm{\delta}^{T}$} & & 
		\multicolumn{1}{l}{$\bm{\delta}^{n} = \bm{\delta}^{T}$} & 
		\multicolumn{1}{l}{$\bm{\delta}^{n} < \bm{\delta}^{T}$} &
		\multicolumn{1}{l}{$\bm{\delta}^{n} > \bm{\delta}^{T}$}\\
		\midrule
		1.1 & 0.001 &   & 0.000 & 0.000 & 0.000 &   & 0.000 & 0.000 & 0.000 \\ 
		1.2 & 0.004 &   & 0.000 & 0.000 & 0.000 &   & 0.000 & 0.000 & 0.000 \\ 
		1.3 & 0.002 &   & 0.000 & 0.000 & 0.000 &   & 0.000 & 0.000 & 0.000 \\ 
		\multicolumn{10}{c}{ }\\
		2.1 & 0.047 &   & 0.049 & 0.047 & 0.052 &   & 0.053 & 0.045 & 0.051 \\ 
		2.2 & 0.034 &   & 0.056 & 0.050 & 0.046 &   & 0.046 & 0.047 & 0.046 \\ 
		2.3 & 0.031 &   & 0.049 & 0.046 & 0.056 &   & 0.051 & 0.054 & 0.050 \\ 
		\multicolumn{10}{c}{ }\\
		3.1 & 0.990 &   & 0.807 & 1.000 & 0.366 &   & 0.796 & 1.000 & 0.389 \\ 
		3.2 & 0.934 &   & 0.806 & 1.000 & 0.357 &   & 0.795 & 1.000 & 0.367 \\ 
		3.3 & 0.825 &   & 0.800 & 1.000 & 0.345 &   & 0.804 & 1.000 & 0.343 \\ 
		\multicolumn{10}{c}{ }\\
		4.1 & 1.000 &   & 0.811 & 1.000 & 0.545 &   & 0.832 & 1.000 & 0.615 \\ 
		4.2 & 1.000 &   & 0.813 & 0.999 & 0.520 &   & 0.810 & 1.000 & 0.569 \\ 
		4.3 & 1.000 &   & 0.804 & 1.000 & 0.514 &   & 0.806 & 0.999 & 0.534 \\ 
		\multicolumn{10}{c}{ }\\
		5.1 & 1.000 &   & 0.881 & 0.975 & 0.837 &   & 0.925 & 0.982 & 0.899 \\ 
		5.2 & 1.000 &   & 0.831 & 0.967 & 0.693 &   & 0.871 & 0.967 & 0.790 \\ 
		5.3 & 1.000 &   & 0.809 & 0.958 & 0.696 &   & 0.842 & 0.969 & 0.753 \\ 
		\multicolumn{10}{c}{ }\\
		6.1 & 1.000 &   & 0.824 & 1.000 & 0.552 &   & 0.835 & 1.000 & 0.633 \\ 
		6.2 & 1.000 &   & 0.805 & 1.000 & 0.512 &   & 0.819 & 0.999 & 0.564 \\ 
		6.3 & 1.000 &   & 0.801 & 0.999 & 0.514 &   & 0.799 & 1.000 & 0.541 \\ 
		\multicolumn{10}{c}{ }\\
		7.1 & 0.007 &   & 0.000 & 0.000 & 0.000 &   & 0.000 & 0.000 & 0.000 \\ 
		7.2 & 0.010 &   & 0.000 & 0.000 & 0.000 &   & 0.000 & 0.000 & 0.000 \\ 
		7.3 & 0.007 &   & 0.000 & 0.000 & 0.000 &   & 0.000 & 0.000 & 0.000 \\ 
		\multicolumn{10}{c}{ }\\
		8.1 & 1.000 &   & 0.808 & 1.000 & 0.494 &   & 0.811 & 1.000 & 0.538 \\ 
		8.2 & 1.000 &   & 0.804 & 1.000 & 0.464 &   & 0.808 & 1.000 & 0.497 \\ 
		8.3 & 1.000 &   & 0.805 & 1.000 & 0.461 &   & 0.802 & 1.000 & 0.471 \\ 
		\bottomrule
	\end{tabular}
}
\end{table}

% latex table generated in R 3.5.3 by xtable 1.8-3 package
% Mon Feb 03 07:06:42 2020
\begin{table}[htbp]
	\small\sf\centering
	\caption{Average sample size to correctly conclude superiority for different trial designs (AD = adaptive design, FD = fixed design, SD = group sequential design) and anticipated treatment differences ($\bm{\delta}^{n}$) after applying the Compensatory decision rule with equal weights. 
	Data generating mechanisms with a hyphen should not result in treatment superiority.} 
	\label{tab:CompareDesigns_nStop}
	\resizebox{\textwidth}{!}{
		\begin{tabular}{lrp{0.01\textwidth}rrrp{0.01\textwidth}rrr}
		\toprule
		DGM & \multicolumn{2}{l}{AD} &
		\multicolumn{4}{l}{FD} &
		\multicolumn{3}{l}{SD}\\
		& & & 
		\multicolumn{1}{l}{$\bm{\delta}^{n} = \bm{\delta}^{T}$} & 
		\multicolumn{1}{l}{$\bm{\delta}^{n} < \bm{\delta}^{T}$} &
		\multicolumn{1}{l}{$\bm{\delta}^{n} > \bm{\delta}^{T}$} & & 
		\multicolumn{1}{l}{$\bm{\delta}^{n} = \bm{\delta}^{T}$} & 
		\multicolumn{1}{l}{$\bm{\delta}^{n} < \bm{\delta}^{T}$} &
		\multicolumn{1}{l}{$\bm{\delta}^{n} > \bm{\delta}^{T}$}\\
		\midrule
		1.1 & - &   & - & - & - &   & - & - & - \\ 
		1.2 & - &   & - & - & - &   & - & - & - \\ 
		1.3 & - &   & - & - & - &   & - & - & - \\ 
		\multicolumn{10}{c}{ }\\
		2.1 & - &   & - & - & - &   & - & - & - \\ 
		2.2 & - &   & - & - & - &   & - & - & - \\ 
		2.3 & - &   & - & - & - &   & - & - & - \\ 
		\multicolumn{10}{c}{ }\\
		3.1 & 159 &   & 108 & 1000 & 26 &   & 91 & 354 & 25 \\ 
		3.2 & 211 &   & 154 & 1000 & 38 &   & 130 & 403 & 37 \\ 
		3.3 & 243 &   & 199 & 1000 & 49 &   & 169 & 452 & 48 \\ 
		\multicolumn{10}{c}{ }\\
		4.1 & 39 &   & 26 & 108 & 11 &   & 20 & 53 & 9 \\ 
		4.2 & 60 &   & 38 & 154 & 16 &   & 31 & 77 & 15 \\ 
		4.3 & 80 &   & 49 & 199 & 21 &   & 41 & 101 & 20 \\ 
		\multicolumn{10}{c}{ }\\
		5.1 & 9 &   & 6 & 11 & 4 &   & 4 & 6 & 3 \\ 
		5.2 & 14 &   & 9 & 16 & 5 &   & 7 & 10 & 4 \\ 
		5.3 & 18 &   & 11 & 21 & 7 &   & 8 & 14 & 6 \\ 
		\multicolumn{10}{c}{ }\\
		6.1 & 37 &   & 25 & 103 & 11 &   & 19 & 50 & 9 \\ 
		6.2 & 57 &   & 36 & 147 & 15 &   & 30 & 74 & 14 \\ 
		6.3 & 76 &   & 47 & 191 & 20 &   & 39 & 96 & 19 \\ 
		\multicolumn{10}{c}{ }\\
		7.1 & - &   & - & - & - &   & - & - & - \\ 
		7.2 & - &   & - & - & - &   & - & - & - \\ 
		7.3 & - &   & - & - & - &   & - & - & - \\ 
		\multicolumn{10}{c}{ }\\
		8.1 & 62 &   & 41 & 298 & 15 &   & 34 & 114 & 13 \\ 
		8.2 & 94 &   & 59 & 426 & 22 &   & 49 & 166 & 21 \\ 
		8.3 & 123 &   & 76 & 553 & 28 &   & 65 & 214 & 27 \\ 
		\bottomrule
	\end{tabular}
}
\end{table}

% latex table generated in R 3.5.3 by xtable 1.8-3 package
% Mon Feb 03 08:14:28 2020
\begin{table}[htbp]
	\small\sf\centering
	\caption{Average bias for different trial designs (AD = adaptive design, FD = fixed design, SD = group sequential design) and anticipated treatment differences ($\bm{\delta}^{n}$) after applying the Compensatory decision rule with equal weights.} 
	\label{tab:CompareDesigns_bias}
	\resizebox{\textwidth}{!}{
		\begin{tabular}{lrp{0.01\textwidth}rrrp{0.01\textwidth}rrr}
		\toprule
		DGM & \multicolumn{2}{l}{AD} &
		\multicolumn{4}{l}{FD} &
		\multicolumn{3}{l}{SD}\\
		& & & 
		\multicolumn{1}{l}{$\bm{\delta}^{n} = \bm{\delta}^{T}$} & 
		\multicolumn{1}{l}{$\bm{\delta}^{n} < \bm{\delta}^{T}$} &
		\multicolumn{1}{l}{$\bm{\delta}^{n} > \bm{\delta}^{T}$} & & 
		\multicolumn{1}{l}{$\bm{\delta}^{n} = \bm{\delta}^{T}$} & 
		\multicolumn{1}{l}{$\bm{\delta}^{n} < \bm{\delta}^{T}$} &
		\multicolumn{1}{l}{$\bm{\delta}^{n} > \bm{\delta}^{T}$}\\
		\midrule
		1.1 & (0.00, 0.00) &  & (0.00, 0.00) & (0.00, 0.00) & (-0.00, 0.00) &  & (0.00, 0.00) & (0.00, 0.00) & (0.00, 0.00) \\ 
		1.2 & (0.00, 0.00) &  & (0.00, 0.00) & (0.00, 0.00) & (-0.00, -0.00) &  & (0.00, 0.00) & (0.00, 0.00) & (0.00, 0.00) \\ 
		1.3 & (0.00, 0.00) &  & (0.00, 0.00) & (0.00, 0.00) & (0.00, -0.00) &  & (0.00, 0.00) & (0.00, 0.00) & (0.00, 0.00) \\ 
		\multicolumn{10}{c}{ }\\
		2.1 & (0.01, 0.01) &  & (-0.00, -0.00) & (0.00, 0.00) & (-0.00, 0.00) &  & (0.00, -0.00) & (0.00, -0.00) & (-0.00, 0.00) \\ 
		2.2 & (0.01, 0.01) &  & (0.00, 0.00) & (0.00, -0.00) & (0.00, 0.00) &  & (0.00, 0.00) & (0.00, 0.00) & (-0.00, 0.00) \\ 
		2.3 & (0.01, 0.01) &  & (-0.00, -0.00) & (0.00, 0.00) & (0.00, 0.00) &  & (0.00, 0.00) & (0.00, -0.00) & (0.00, 0.00) \\ 
		\multicolumn{10}{c}{ }\\
		3.1 & (0.05, 0.05) &  & (0.00, 0.00) & (0.00, 0.00) & (0.00, 0.00) &  & (0.01, 0.01) & (0.00, 0.00) & (0.02, 0.01) \\ 
		3.2 & (0.05, 0.05) &  & (0.00, 0.00) & (0.00, 0.00) & (0.00, -0.00) &  & (0.01, 0.01) & (0.00, 0.00) & (0.01, 0.01) \\ 
		3.3 & (0.05, 0.05) &  & (0.00, 0.00) & (0.00, 0.00) & (-0.00, -0.00) &  & (0.01, 0.01) & (0.01, 0.01) & (0.01, 0.01) \\ 
		\multicolumn{10}{c}{ }\\
		4.1 & (0.07, 0.07) &  & (0.00, 0.00) & (0.00, 0.00) & (0.00, -0.00) &  & (0.03, 0.03) & (0.01, 0.01) & (0.04, 0.05) \\ 
		4.2 & (0.07, 0.08) &  & (0.00, 0.00) & (0.00, 0.00) & (-0.00, -0.00) &  & (0.03, 0.03) & (0.01, 0.01) & (0.04, 0.04) \\ 
		4.3 & (0.08, 0.08) &  & (0.00, 0.00) & (0.00, 0.00) & (-0.00, 0.00) &  & (0.02, 0.02) & (0.01, 0.01) & (0.02, 0.03) \\ 
		\multicolumn{10}{c}{ }\\
		5.1 & (0.04, 0.04) &  & (0.00, -0.01) & (0.00, 0.00) & (-0.01, 0.00) &  & (0.03, 0.04) & (0.04, 0.04) & (0.06, 0.06) \\ 
		5.2 & (0.07, 0.07) &  & (-0.01, -0.01) & (0.00, 0.01) & (-0.00, 0.00) &  & (0.05, 0.04) & (0.04, 0.04) & (0.08, 0.08) \\ 
		5.3 & (0.09, 0.09) &  & (0.00, -0.01) & (0.00, 0.00) & (0.00, 0.00) &  & (0.06, 0.06) & (0.04, 0.04) & (0.07, 0.07) \\ 
		\multicolumn{10}{c}{ }\\
		6.1 & (0.06, 0.07) &  & (0.00, 0.00) & (0.00, -0.00) & (-0.01, 0.01) &  & (0.03, 0.03) & (0.01, 0.02) & (0.04, 0.06) \\ 
		6.2 & (0.07, 0.07) &  & (0.00, -0.00) & (0.00, -0.00) & (-0.00, -0.00) &  & (0.02, 0.03) & (0.02, 0.02) & (0.04, 0.03) \\ 
		6.3 & (0.08, 0.08) &  & (0.00, -0.00) & (0.00, -0.00) & (0.00, -0.00) &  & (0.02, 0.02) & (0.01, 0.01) & (0.02, 0.03) \\ 
		\multicolumn{10}{c}{ }\\
		7.1 & (0.00, 0.00) &  & (0.00, 0.00) & (0.00, 0.00) & (0.00, 0.00) &  & (0.00, 0.00) & (0.00, 0.00) & (0.00, 0.00) \\ 
		7.2 & (0.01, 0.00) &  & (0.00, 0.00) & (0.00, 0.00) & (-0.00, -0.00) &  & (0.00, 0.00) & (0.00, 0.00) & (0.00, 0.00) \\ 
		7.3 & (0.00, 0.00) &  & (0.00, 0.00) & (0.00, 0.00) & (-0.00, -0.00) &  & (0.00, 0.00) & (0.00, 0.00) & (0.00, 0.00) \\ 
		\multicolumn{10}{c}{ }\\
		8.1 & (0.07, 0.06) &  & (0.00, 0.00) & (0.00, 0.00) & (0.00, 0.00) &  & (0.02, 0.02) & (0.00, 0.00) & (0.02, 0.03) \\ 
		8.2 & (0.07, 0.07) &  & (0.00, 0.00) & (0.00, 0.00) & (0.00, -0.00) &  & (0.02, 0.02) & (0.01, 0.00) & (0.02, 0.02) \\ 
		8.3 & (0.07, 0.07) &  & (0.00, 0.00) & (0.00, 0.00) & (0.00, 0.00) &  & (0.02, 0.02) & (0.00, 0.00) & (0.02, 0.02) \\ 
		\bottomrule
	\end{tabular}
}
\end{table}

\subsection{Discussion}
Each of the designs is compatible with the proposed multivariate decision-making framework and has specific advantages.
Although fixed designs perform well under accurate sample size estimation, a priori sample sample size estimation is difficult when multiple parameters are unknown. 
Sequential or adaptive designs may be beneficial to deal with this parameter uncertainty, albeit at the expense of bias.
Whereas adaptive designs deal most flexibly with parameter uncertainty, group sequential designs limit bias more than adaptive designs.

We remark that adaptive designs in particular have their practical challenges.
First, updating adaptive designs might require a large logistic effort, which increases with the size of the study. 
These designs are therefore easier to implement in small phase I or II studies compared to confirmatory phase II or III studies.
Second, we find that the current literature does not offer clear guidance on the specification of adaptive design parameters.
Further elaboration on the choice of these parameters would undoubtedly serve trials that stop data collection adaptively. 

%%%%%%%%%%%%%%%%%%%%%%%%%%%%%%%%%%%%%%%%%%%%%%%%%%%%%%%%%%%%%%%%%%%%%%%%%%%%%
%																			%
%			APPENDIX E Comparison of prior specifications					%
%																			%
%%%%%%%%%%%%%%%%%%%%%%%%%%%%%%%%%%%%%%%%%%%%%%%%%%%%%%%%%%%%%%%%%%%%%%%%%%%%%
\section{Numerical evaluation: Comparison of prior specifications}\label{app:compare_priors}
To demonstrate the influence of prior information on the performance of the Compensatory decision rule, we specified six different sets of prior hyperparameters, that are presented in Table \ref{tab:priors}.
Two of these prior specifications are assumed to be uninformative ($1-2$).
Prior $1$ is the non-informative prior that we used in the \textit{\nameref{sec:evaluation}} Section and serves as a reference prior in this comparison.
Prior $2$ is Jeffreys' prior, which is well-known for its property to remain invariant under transformation of parameters \cite{Yang1996}.
This is useful since our main interest is typically in the transformed parameters $\bm{\delta}$ rather than the marginal probabilities $\bm{\theta}_{j}$ or the cell probabilities $\bm{\phi}_{j}$, on which the treatment-specific prior distributions are specified.
Four informative prior specifications (priors $3-6$) include $20$ additional observations (i.e. $n^{0}_{j}=20$).
These $20$ additional observations show the effects on decisions when the number of prior observations is either higher or lower than sample size $n_{j}$.
The former occurs in data generating mechanisms with large treatment differences that require sample sizes smaller than $20$ (e.g. treatment difference $5$), while the latter occurs when treatment differences are small and sample sizes are larger (e.g. treatment difference $3$). 
Prior specifications $3-6$ differ on the correspondence between the prior treatment difference $\bm{\delta}^{0}$ and the true treatment difference $\bm{\delta}^{T}$ used for data generation.
Specifically, we included prior differences that are identical (prior $3$), smaller (prior $4$), larger (prior $5$) or opposite (prior $6$) to the true treatment difference.
The bivariate beta distributions of prior specifications $1$, $2$, and $3$ for data generating mechanism $2.2$ are visually presented in Figure \ref{fig:bibeta}.
We ran the introduced procedure for a fixed design (Steps $2$ and $3$ of Algorithm \ref{alg:fixed}), using sample size $n_{j}$ for a fixed design estimated based on true treatment differences. 
These sample sizes were also used in Section \textit{\nameref{sec:evaluation}} and presented in Table \ref{tab:CompareRules_nStop}.

\begin{table}[htbp]
	\small\sf\centering
	\caption{Prior specifications used for numerical evaluation. Prior hyperparameters $\bm{\alpha}^{0}_{j}=n^{0}_{j}\bm{\phi}^{0}_{j}$. True parameters $\bm{\phi}^{T}_{E}$, $\bm{\phi}^{T}_{C}$ and $\bm{\delta}^{T}$ can be obtained via the simulation conditions presented in Table \ref{tab:conditions}.} 
	\label{tab:priors}
	\begin{tabular}{llllll}
		\toprule
		Prior & $n^{0}_{j}$ & $\bm{\phi}^{0}_{E}$ & $\bm{\phi}^{0}_{C}$ & $\bm{\delta}^{0}$ &  \\
		\midrule
		
		1 & $\frac{1}{25}$ & $\frac{1}{4}$ & $\frac{1}{4}$ & $0$ & \\
		2 & $2$ & $\frac{1}{4}$ & $\frac{1}{4}$ & $0$ & \\
		3 & $20$ & $\bm{\phi}^{T}_{E} $ & $\bm{\phi}^{T}_{C} $ & $\bm{\delta}^{T}$ & \\
		4 & $20$ & $\bm{\phi}^{T}_{E} - 0.05$ & $\bm{\phi}^{T}_{C} + 0.05$ & $\bm{\delta}^{T} - 0.10$ & \\
		5 & $20$ & $\bm{\phi}^{T}_{E} + 0.05$ & $\bm{\phi}^{T}_{C} - 0.05$ & $\bm{\delta}^{T} + 0.10$ & \\
		6 & $20$ & $\bm{\phi}^{T}_{C}$ & $\bm{\phi}^{T}_{E}$ & $- \bm{\delta}^{T}$ & \\
		\bottomrule
	\end{tabular}
\end{table}

\subsection{Results}
The two uninformative priors do not noticeably influence the probability to conclude superiority (Table \ref{tab:ComparePriors_pSup}) or the average treatment effect (Table \ref{tab:ComparePriors_bias}) in the majority of data generating mechanisms. 
An exception is a large treatment difference ($5.1-5.3$) where Jeffreys' prior (i.e. prior $2$) lowered power and biased the treatment effect downwards. 
Here, $\bm{s}_{j}$ is too small to satisfy $\bm{\alpha}^{n}_{j} \approx \bm{s}_{j}$.
A smaller prior sample size $n^{0}_{j}$ (prior $1$) resulted in an unbiased estimate of the average treatment difference.

An informative prior distribution influences the probability to conclude superiority as well as the average treatment effect, depending on prior treatment difference $\bm{\delta}^{0}$. 
Prior information improves decision-making when the prior treatment effect equals the true treatment effect (i.e. $\bm{\delta}^{0} = \bm{\delta}^{T}$; prior $3$).
This situation increases power, without influencing Type I error or the average posterior treatment effect

In contrast, prior information affects the decision when prior and true treatment effects do not correspond. 
When the prior treatment effect is less strong than the true treatment effect (i.e. $\bm{\delta}^{0}<\bm{\delta}^{T}$, prior $4$), the Type I error as well as the probability to conclude superiority correctly are lowered, although the effect is masked in treatment differences $4-6$ by the (relatively large) number of prior observations $n^{0}_{j}$. 
Moreover, the average posterior treatment effect is lower than the treatment effect of the data $\bm{\delta}^{T}$ especially when the treatment difference is large and the required sample size is low ($5.1-5.3$). 

When the prior treatment effect is stronger than the true treatment effect (i.e. $\bm{\delta}^{0}>\bm{\delta}^{T}$, prior $5$), the Type I error and the probability to conclude superiority are above the planned $.05$ and $.80$. 
Moreover, the average posterior treatment effect is exceeds the treatment effect of the data $\bm{\delta}^{T}$, in particular when the treatment effect is large ($5.1-5.3$). 
An opposite prior treatment effect (i.e. $\bm{\delta}^{0} = - \bm{\delta}^{T}$, prior $6$) results in a lower probability to conclude superiority as well as an average posterior treatment effect that differs from true treatment difference $\bm{\delta}^{T}$.

In general, the effect of prior information is strongest in $5.1-5.3$ and $8.1-8.3$, where prior sample size $n_{j}^{0}$ is relatively large compared to sample size $n_{j}$, resulting in a larger influence of $\bm{\alpha}^{0}_{j}$ on $\bm{\alpha}^{n}_{j}$.
Note that in practice, a difference between results with and without prior information signals a conflict between the informative prior and the data, and does not necessarily reflect an invalid decision.

% latex table generated in R 3.5.3 by xtable 1.8-3 package
% Mon Feb 03 07:11:05 2020
\begin{table}[htbp]
	\small\sf\centering
	\caption{P(Conclude superiority) for six different prior specifications (see Table \ref{tab:priors}) after applying the Compensatory decision rule with equal weights.} 
	\label{tab:ComparePriors_pSup}
	\begin{tabular}{lrrrrrr}
		\toprule
		DGM &
		\multicolumn{1}{l}{1} &
		\multicolumn{1}{l}{2} & 
		\multicolumn{1}{l}{3} &
		\multicolumn{1}{l}{4} & 
		\multicolumn{1}{l}{5} & 
		\multicolumn{1}{l}{6}\\
		\midrule
		1.1 & 0.000 & 0.000 & 0.000 & 0.000 & 0.000 & 0.000 \\ 
		1.2 & 0.000 & 0.000 & 0.000 & 0.000 & 0.000 & 0.000 \\ 
		1.3 & 0.000 & 0.000 & 0.000 & 0.000 & 0.000 & 0.000 \\ 
		\multicolumn{7}{c}{ }\\
		2.1 & 0.049 & 0.055 & 0.049 & 0.037 & 0.067 & 0.049 \\ 
		2.2 & 0.056 & 0.050 & 0.049 & 0.038 & 0.054 & 0.055 \\ 
		2.3 & 0.049 & 0.048 & 0.045 & 0.035 & 0.064 & 0.053 \\ 
		\multicolumn{7}{c}{ }\\
		3.1 & 0.807 & 0.798 & 0.872 & 0.753 & 0.951 & 0.606 \\ 
		3.2 & 0.806 & 0.803 & 0.855 & 0.775 & 0.911 & 0.674 \\ 
		3.3 & 0.800 & 0.793 & 0.841 & 0.782 & 0.895 & 0.688 \\ 
		\multicolumn{7}{c}{ }\\
		4.1 & 0.811 & 0.791 & 0.987 & 0.905 & 0.999 & 0.043 \\ 
		4.2 & 0.813 & 0.794 & 0.967 & 0.867 & 0.990 & 0.178 \\ 
		4.3 & 0.804 & 0.802 & 0.939 & 0.854 & 0.979 & 0.313 \\ 
		\multicolumn{7}{c}{ }\\
		5.1 & 0.881 & 0.704 & 1.000 & 1.000 & 1.000 & 0.000 \\ 
		5.2 & 0.831 & 0.787 & 1.000 & 1.000 & 1.000 & 0.000 \\ 
		5.3 & 0.809 & 0.761 & 1.000 & 0.998 & 1.000 & 0.000 \\ 
		\multicolumn{7}{c}{ }\\
		6.1 & 0.824 & 0.789 & 0.990 & 0.900 & 0.999 & 0.030 \\ 
		6.2 & 0.805 & 0.797 & 0.965 & 0.874 & 0.991 & 0.145 \\ 
		6.3 & 0.801 & 0.792 & 0.946 & 0.856 & 0.984 & 0.282 \\ 
		\multicolumn{7}{c}{ }\\
		7.1 & 0.000 & 0.000 & 0.000 & 0.000 & 0.000 & 0.000 \\ 
		7.2 & 0.000 & 0.000 & 0.000 & 0.000 & 0.000 & 0.000 \\ 
		7.3 & 0.000 & 0.000 & 0.000 & 0.000 & 0.000 & 0.000 \\ 
		\multicolumn{7}{c}{ }\\
		8.1 & 0.808 & 0.792 & 0.957 & 0.837 & 0.992 & 0.224 \\ 
		8.2 & 0.804 & 0.799 & 0.925 & 0.823 & 0.975 & 0.387 \\ 
		8.3 & 0.805 & 0.793 & 0.896 & 0.811 & 0.955 & 0.480 \\ 
		\bottomrule
	\end{tabular}
\end{table}

% latex table generated in R 3.5.3 by xtable 1.8-3 package
% Mon Feb 03 07:20:52 2020
\begin{table}[htbp]
	\small\sf\centering
	\caption{Average bias for six different prior specifications (see Table \ref{tab:priors}) after applying the Compensatory decision rule with equal weights.} 
	\label{tab:ComparePriors_bias}
	\resizebox{\textwidth}{!}{
		\begin{tabular}{lrrrrrr}
		\toprule
		DGM &
		\multicolumn{1}{l}{1} &
		\multicolumn{1}{l}{2} & 
		\multicolumn{1}{l}{3} &
		\multicolumn{1}{l}{4} & 
		\multicolumn{1}{l}{5} & 
		\multicolumn{1}{l}{6}\\
		\midrule
		1.1 & (0.00, 0.00) & (0.00, 0.00) & (0.00, 0.00) & (0.00, 0.00) & (0.00, 0.00) & (0.01, 0.01) \\ 
		1.2 & (0.00, 0.00) & (0.00, 0.00) & (0.00, 0.00) & (0.00, 0.00) & (0.00, 0.00) & (0.01, 0.01) \\ 
		1.3 & (0.00, 0.00) & (0.00, 0.00) & (0.00, 0.00) & (0.00, 0.00) & (0.00, 0.00) & (0.01, 0.01) \\ 
		\multicolumn{7}{l}{ }\\
		2.1 & (-0.00, -0.00) & (0.00, 0.00) & (-0.00, 0.00) & (-0.00, -0.00) & (0.00, 0.00) & (0.00, -0.00) \\ 
		2.2 & (0.00, 0.00) & (-0.00, -0.00) & (-0.00, -0.00) & (-0.00, -0.00) & (0.00, 0.00) & (0.00, -0.00) \\ 
		2.3 & (-0.00, -0.00) & (0.00, -0.00) & (-0.00, -0.00) & (-0.00, -0.00) & (0.00, 0.00) & (-0.00, -0.00) \\ 
		\multicolumn{7}{l}{ }\\
		3.1 & (0.00, 0.00) & (0.00, 0.00) & (0.00, 0.00) & (-0.02, -0.02) & (0.01, 0.02) & (-0.03, -0.03) \\ 
		3.2 & (0.00, 0.00) & (0.00, 0.00) & (0.00, 0.00) & (-0.01, -0.01) & (0.01, 0.01) & (-0.02, -0.02) \\ 
		3.3 & (0.00, 0.00) & (0.00, 0.00) & (0.00, 0.00) & (-0.01, -0.01) & (0.01, 0.01) & (-0.02, -0.02) \\ 
		\multicolumn{7}{l}{ }\\
		4.1 & (0.00, 0.00) & (-0.01, -0.02) & (0.00, 0.00) & (-0.04, -0.04) & (0.04, 0.04) & (-0.17, -0.17) \\ 
		4.2 & (0.00, 0.00) & (-0.01, -0.01) & (0.00, 0.00) & (-0.03, -0.04) & (0.04, 0.03) & (-0.14, -0.14) \\ 
		4.3 & (0.00, 0.00) & (-0.01, -0.01) & (0.00, 0.00) & (-0.03, -0.03) & (0.03, 0.03) & (-0.11, -0.12) \\ 
		\multicolumn{7}{l}{ }\\
		5.1 & (0.00, -0.01) & (-0.10, -0.10) & (0.00, 0.00) & (-0.08, -0.08) & (0.08, 0.08) & (-0.62, -0.62) \\ 
		5.2 & (-0.01, -0.01) & (-0.07, -0.07) & (0.00, 0.00) & (-0.07, -0.07) & (0.07, 0.07) & (-0.55, -0.55) \\ 
		5.3 & (0.00, -0.01) & (-0.06, -0.06) & (0.00, 0.00) & (-0.06, -0.06) & (0.06, 0.07) & (-0.52, -0.52) \\ 
		\multicolumn{7}{l}{ }\\
		6.1 & (0.00, 0.00) & (-0.03, 0.00) & (0.00, -0.00) & (-0.05, -0.05) & (0.04, 0.04) & (-0.36, 0.00) \\ 
		6.2 & (0.00, -0.00) & (-0.02, 0.00) & (0.00, -0.00) & (-0.04, -0.03) & (0.03, 0.04) & (-0.29, -0.00) \\ 
		6.3 & (0.00, -0.00) & (-0.02, -0.00) & (0.00, 0.00) & (-0.03, -0.03) & (0.03, 0.03) & (-0.24, -0.00) \\ 
		\multicolumn{7}{l}{ }\\
		7.1 & (0.00, 0.00) & (0.00, 0.00) & (0.00, 0.00) & (0.00, 0.00) & (0.00, 0.00) & (-0.01, 0.02) \\ 
		7.2 & (0.00, 0.00) & (0.00, 0.00) & (0.00, 0.00) & (0.00, 0.00) & (0.00, 0.00) & (-0.01, 0.02) \\ 
		7.3 & (0.00, 0.00) & (0.00, 0.00) & (0.00, 0.00) & (0.00, 0.00) & (0.00, 0.00) & (-0.01, 0.02) \\ 
		\multicolumn{7}{l}{ }\\
		8.1 & (0.00, 0.00) & (-0.01, 0.00) & (0.00, 0.00) & (-0.03, -0.03) & (0.03, 0.03) & (-0.16, -0.05) \\ 
		8.2 & (0.00, 0.00) & (-0.01, 0.00) & (0.00, 0.00) & (-0.03, -0.02) & (0.03, 0.03) & (-0.12, -0.04) \\ 
		8.3 & (0.00, 0.00) & (-0.01, 0.00) & (0.00, 0.00) & (-0.02, -0.02) & (0.02, 0.02) & (-0.10, -0.03) \\ 
		\bottomrule
	\end{tabular}
}
\end{table}

\end{appendices}

\end{document}